\documentclass[12pt]{article}
\usepackage{fullpage, doublespace}
\usepackage{amssymb, amsthm, amsmath}
\usepackage{graphicx}
\usepackage{authblk}
\usepackage[authoryear]{natbib}
\usepackage{bm}
\usepackage{lineno}
\usepackage{times}
\usepackage{float}
\usepackage{soul}
\usepackage{dutchcal} 
\usepackage{adjustbox}
\usepackage{makecell}
\usepackage[skip=0.1\baselineskip]{caption}
\usepackage[ruled]{algorithm2e}

\newcommand*\patchAmsMathEnvironmentForLineno[1]{%
  \expandafter\let\csname old#1\expandafter\endcsname\csname #1\endcsname
  \expandafter\let\csname oldend#1\expandafter\endcsname\csname end#1\endcsname
  \renewenvironment{#1}%
     {\linenomath\csname old#1\endcsname}%
     {\csname oldend#1\endcsname\endlinenomath}}%
\newcommand*\patchBothAmsMathEnvironmentsForLineno[1]{%
  \patchAmsMathEnvironmentForLineno{#1}%
  \patchAmsMathEnvironmentForLineno{#1*}}%
\AtBeginDocument{%
\patchBothAmsMathEnvironmentsForLineno{equation}%
\patchBothAmsMathEnvironmentsForLineno{align}%
\patchBothAmsMathEnvironmentsForLineno{flalign}%
\patchBothAmsMathEnvironmentsForLineno{alignat}%
\patchBothAmsMathEnvironmentsForLineno{gather}%
\patchBothAmsMathEnvironmentsForLineno{multline}%
}

\newcommand{\muxsi}{\mu_X(\bs_i, t)}

\newcommand{\muysi}{\mu_Y(\bs_i, t)}

\newcommand{\muysk}{\mu_Y(\bs_k, t)}

\newcommand{\sxxsisi}{\sigma_{XX}(\bs_i, \bs_i;t)}

\newcommand{\sxxsisk}{\sigma_{XX}(\bs_i, \bs_k;t)}

\newcommand{\syysisi}{\sigma_{YY}(\bs_i, \bs_i;t)}

\newcommand{\syysisk}{\sigma_{YY}(\bs_i, \bs_k;t)}

\newcommand{\sxysisi}{\sigma_{XY}(\bs_i, \bs_i;t)}

\newcommand{\sxysisk}{\sigma_{XY}(\bs_i, \bs_k;t)}

\newcommand{\betasnsi}{\frac{\beta(\bs_i)}{N(\bs_i)}}
\newcommand{\phinsi}{\frac{\phi}{N(\bs_i)}}
\newcommand{\Xsi}{X(\bs_i,t)}

\newcommand{\Ysi}{Y(\bs_i,t)}

\newcommand{\Ysk}{Y(\bs_k,t)}

\newcommand{\skinnsi}{\bs_k \in \mathcal{N}(\bs_i)}

\newcommand{\btheta}{ \mbox{\boldmath $\theta$}}
\newcommand{\bmu}{ \mbox{\boldmath $\mu$}}
\newcommand{\balpha}{ \mbox{\boldmath $\alpha$}}
\newcommand{\bbeta}{ \mbox{\boldmath $\beta$}}

\newcommand{\bdelta}{ \mbox{\boldmath $\delta$}}
\newcommand{\bzeta}{ \mbox{\boldmath $\zeta$}}
\newcommand{\blambda}{ \mbox{\boldmath $\lambda$}}
\newcommand{\bgamma}{ \mbox{\boldmath $\gamma$}}

\newcommand{\bDelta}{ \mbox{\boldmath $\Delta$}}

\newcommand{\bGamma}{ \mbox{\boldmath $\Gamma$}}

\newcommand{\ba}{ \mbox{\bf a}}

\newcommand{\by}{ \mbox{\bf y}}

\newcommand{\bs}{ \mbox{\bf s}}

\newcommand{\beq}{ \begin{equation}}
\newcommand{\eeq}{ \end{equation}}
\newcommand{\beqn}{ \begin{eqnarray}}
\newcommand{\eeqn}{ \end{eqnarray}}

\title{\vspace{-2cm}A Gaussian-process approximation to a spatial SIR process using moment closures and emulators}
\author[1]{Parker Trostle}
\author[2]{Joseph Guinness}
\author[1]{Brian J. Reich}
\affil[1]{Department of Statistics, North Carolina State University}
\affil[2]{Department of Statistics and Data Science, Cornell University}
\date{}

\begin{document}

\maketitle

\begin{abstract}\begin{singlespace}
\noindent The dynamics that govern disease spread are hard to model because infections are functions of both the underlying pathogen as well as human or animal behavior. This challenge is increased when modeling how diseases spread between different spatial locations. Many proposed spatial epidemiological models require trade-offs to fit, either by abstracting away theoretical spread dynamics, fitting a deterministic model, or by requiring large computational resources for many simulations. We propose an approach that approximates the complex spatial spread dynamics with a Gaussian process. We first propose a flexible spatial extension to the well-known SIR stochastic process, and then we derive a moment-closure approximation to this stochastic process. This moment-closure approximation yields ordinary differential equations for the evolution of the means and covariances of the susceptibles and infectious through time. Because these ODEs are a bottleneck to fitting our model by MCMC, we approximate them using a low-rank emulator. This approximation serves as the basis for our hierarchical model for noisy, underreported counts of new infections by spatial location and time. We demonstrate using our model to conduct inference on simulated infections from the underlying, true spatial SIR jump process. We then apply our method to model counts of new Zika infections in Brazil from late 2015 through early 2016.

\vspace{12pt}
{\bf Key words:} Emulator models,  moment-closure approximations, spatiotemporal epidemiology\end{singlespace}\end{abstract}

\section{Introduction}

There have been many approaches proposed for modeling the spatial spread of disease. A common approach is to rely on generalized linear mixed-effects models (GLMMs), which typically use spatial random effects and spatially varying covariates \citep[e.g.,][]{haredashtpolson, arruda2017land, thanapongtharm2014}. The benefit to using GLMMs is software to fit these models is readily available, and the interpretation of the model output is widely understood. However, these models may serve as abstractions from the true underlying disease-spread process. Alternative approaches are often based in theory and are typically a form of compartmental model \citep[e.g.,][]{burgerchowell, paenglee2017, galvis2022cshl, jones2021}.  They either are deterministic models, which may not be flexible enough for variability in real infection data, or stochastic models that require computationally demanding simulation-based methods to fit such as Approximate Bayesian Computing \citep{beaumont2010}.

Our work aims to combine the benefits of the GLMMs with those of the compartmental models. Our method is built upon three components. The first component is a stochastic, spatial susceptible-infectious-recovered (SIR) model that can model the spread of disease both within and between spatial locations. This underlying, theoretical model is similar to that of \cite{paenglee2017} but is more general and is stochastic. However, it is challenging to model data from a stochastic process such as our proposed spatial SIR process because characterizing the moments of a continuous-time Markov process is not simple \citep[e.g., see][on the SIR process]{allen2008}. 

This leads to the second component of our approach, which is to approximate the spatial SIR process with a Gaussian process using ``moment closure.'' This methodology is based on approximating the moments of the complex SIR process with those of a Gaussian process, and it was used in \cite{isham1991} for the non-spatial SIR process. This ``closes'' the dependency on higher-order moments because the higher-order moments in a Gaussian process are fully characterized by its mean and covariance function. Moment-closure approximations were originally proposed in \cite{whittle1957}. Though uncommon today in the statistics literature, moment-closure approximations have been used extensively elsewhere in the natural sciences \citep[e.g.,][]{forgues2019, kuehn2016}. There have been some proposed spatial moment-closure models, particularly in ecology \citep[e.g.,][]{murrell2004, ernstbartol} and for disease spread on networks where nodes are binary-valued \citep[e.g.,][]{sharkeykiss, chenogura}. Our approach is more general and relies on fewer simplying assumptions than some of the above literature.

There is an unfortunate downside to the moment-closure approach. These approximations are characterized by coupled, ordinary differential equations (ODEs). These ODEs are used to calculate the (marginalized over time) means and spatial covariances for the Gaussian-process approximation. This is a computational bottleneck for even a modest number of spatial locations. 

To address this bottleneck, we implement the third component of our methodology: an emulator-based approximation to the moment-closure forward equations. Emulators, also known as surrogate models, are used to approximate the output of computationally intensive models \citep{gramacy2020}. There is a long history of emulators in the statistics literature. Much of the framework for modern emulators can be traced to \cite{kennedyohagan}, an influential paper that proposed a Bayesian approach to calibrating computer models. There are multiple proposed methods building upon their framework \citep{higdon2004, goldsteinrougier2006, josephmelkote, bayarri2007, qianseepersad}. One of the most influential follow-up methodologies was \cite{higdon2008}, which used an SVD-based approach to design emulators. This approach was extended in \cite{hooten2011} and \cite{leeds2014}. Recently \cite{pratola2018} and \cite{gopalanwikle} extended these SVD-based approaches to computational output stored in tensors, an approach we adapt in this paper. Many other approaches to constructing emulators are possible, however \citep[e.g.,][]{gu2018, massoud2019, thakurchakraborty, reichkalendra2012}.

In this paper, we propose a spatiotemporal, Bayesian hierarchical model for noisy counts of new infections reported that are reported in discrete time. We do so by modeling the latent number of susceptibles using a low-rank emulator model based on a moment-closure approximation to our underlying continuous-time spatial SIR process. Our model may be therefore thought of as a so-called physical-statistical model \citep{berliner1996, berliner2003, dowdStatisticalOverviewPerspectives2014}. We make numerous novel contributions by doing so:
\begin{enumerate}
    \item We provide a more general and flexible spatial SIR model than the discrete-space model in \cite{paenglee2017}, and our proposed model is stochastic. We do so by extending the epidemiological moment-closure work of \cite{isham1991} to include a spatial domain.
    \item We develop tensor-based emulators to approximate the coupled ODEs in moment-closure approximations, allowing for their fast approximation.
    \item The combination of moment-closure approximations and emulators yields a new method to model spatial disease spread using the underlying epidemiological dynamics without needing to simulate repeatedly from a stochastic process. We show our model is able to estimate spatially varying covariates from this complex underlying process.
    \item To the best of our knowledge, no one has used an emulator to model the covariance function of a latent process in a physical-statistical model.
\end{enumerate}

\section{Spatial SIR model}
\label{sec:spatialsir}

We begin by describing our spatial extension to the SIR model. We first propose our spatial SIR jump process in Section \ref{sec:spatialsir_jump}, and then we derive its moment-closure approximation and the resulting forward equations in Section \ref{sec:spatialsir_fe}. A background overview for the closed-population SIR model is provided in Appendix \ref{app:bground}.

\subsection{Spatial jump process}
\label{sec:spatialsir_jump}

Let $\mathcal{D}$ be a spatial lattice with $n_s$ spatial coordinates $\bs$. Let $\mathcal{N}(\bs)$ be the set of spatial locations connected to $\bs$. We denote the susceptibles at $\bs$ at time $t$ as $X(\bs, t)$, and we similarly denote the infectious at $\bs$ at time $t$ as $Y(\bs, t)$. Each location has population size $N(\bs)$ that does not vary in time. The number of recovered is therefore determined by $X(\bs, t), Y(\bs, t),$ and $N(\bs)$, so we do not model them directly.

Define the current state at time $t$ as $\pmb{H}(t) = \{X(\bs_1,t), ..., X(\bs_n,t), Y(\bs_1,t), ..., Y(\bs_n,t)\}$. Consider a sufficiently small $\Delta t$ such that only one new infection or recovery at most may occur at any spatial location. Then there are $2 n_s + 1$ possible events in the interval $(t,t + \Delta t)$ for this sufficiently small $\Delta t$: a new infection in one of the $n_s$ locations, a recovery in one of the $n_s$ locations, or no change. For $i\in\{1,...,n_s\}$, let $I^+(\bs_i,t)$ denote the new infection event such that $\pmb{H}(t + \Delta t) = \pmb{H}(t)$ except $X(\bs_i, t + \Delta t) = X(\bs_i, t) - 1$ and $Y(\bs_i, t + \Delta t) = Y(\bs_i, t) + 1$, and let $R^+(\bs_i, t)$ denote the recovery event such that $\pmb{H}(t+\Delta t)=\pmb{H}(t)$ except $Y(\bs_i, t + \Delta t) = Y(\bs_i, t) - 1$. Then our spatial SIR jump process is defined via the following conditional probabilities:
\begin{align}
\begin{split}
    P\left\{I^+(\bs_i,t) | \pmb{H}(t)\right\} &\approx \frac{\beta(\bs_i)\Delta t}{N(\bs_i)} X(\bs_i, t)Y(\bs_i, t) + \frac{\phi\Delta t}{N(\bs_i)}\sum_{\bs_k \in \mathcal{N}(\bs_i)} X(\bs_i, t)Y(\bs_k, t)\\
    P\left\{R^+(\bs_i,t) | \pmb{H}(t)\right\} &\approx \eta Y(\bs_i,t) \Delta t  \label{eq:spatial_jump_example2}
\end{split}
\end{align}
with the probability of no event, $\pmb{H}(t+\Delta)=\pmb{H}(t)$, equal to one minus the sum of the $2 n_s$ probabilities in (\ref{eq:spatial_jump_example2}). These probabilities are controlled by local infection parameters $\beta(\bs_i)$, spatial infection parameters $\phi$, and a recovery parameter $\eta$. Our model differs therefore from the one proposed in \cite{paenglee2017} by being stochastic and not modeling spatial spread as a function of a non-spatially varying $\beta$ and distance.

\subsection{Forward equations for the spatial SIR jump process}
\label{sec:spatialsir_fe}

Working with the spatial SIR jump process directly to model new infection counts is not practical. Therefore, we now describe the moment-closure approximation to this jump process in which we use an approximating Gaussian process. As described in \cite{isham1991}, there are two approaches to derive the governing ODEs for the means and spatial covariances through time, which we call the forward equations. We begin by demonstrating the simpler of the two approaches, which we call the heuristic approach, by deriving the forward equation for the mean number of susceptibles.

As in Section \ref{sec:spatialsir_jump}, consider $\pmb{H}(t)$ and a sufficiently small $\Delta t$ that only the $2 n_s + 1$ transition events described earlier may occur. For a particular $\bs_i$, there can be two events that occur:
\begin{align}
    X(\bs_i, t + \Delta t) = \begin{cases}
    X(\bs_i, t) - 1 \quad &\text{with probability} \quad P\{I^+(\bs_i, t) | \pmb{H}(t)\} \\
    X(\bs_i, t) \quad &\text{with probability} \quad 1 - P\{I^+(\bs_i, t) | \pmb{H}(t)\}
    \end{cases}
\end{align}
where $P\{I^+(\bs_i, t) | \pmb{H}(t)\}$ was defined above in \eqref{eq:spatial_jump_example2}. The expected change in $X(\bs_i, t)$ is therefore $-P\{I^+(\bs_i, t) | \pmb{H}(t)\}$. Taking the expectation of this quantity with respect to $\pmb{H}(t)$ being jointly distributed as a normal random vector and then taking the limit as $\Delta t \rightarrow 0$ yields:
\begin{align}
\begin{split}
    \frac{d\mu_X(\bs_i,t)}{dt} =& -\frac{\beta(\bs_i)}{N(\bs_i)}\bigl(\mu_X(\bs_i,t) \mu_Y(\bs_i,t) + \sigma_{XY}(\bs_i, \bs_i; t)\bigr) \\
    &- \frac{\phi}{N(\bs_i)}\sum_{\bs_k\in\mathcal{N}(\bs_i)}\bigl(\mu_X(\bs_i,t)\mu_Y(\bs_k,t) + \sigma_{XY}(\bs_i,\bs_k; t)\bigr)\label{eq:dmuxst_deriv}
\end{split}
\end{align}
where $\mu_X(\bs_i,t)$ is the mean susceptibles at $\bs_i$ at time $t$, $\mu_Y(\bs_i,t)$ is the mean infectious at $\bs_i$ at time $t$, and $\sigma_{XY}(\bs_i, \bs_k;t)$ is the covariance between the susceptibles at $\bs_i$ and infectious at $\bs_k$ at time $t$. Though not present in \eqref{eq:dmuxst_deriv}, we define $\sigma_{XX}(\bs_i, \bs_k; t)$ and $\sigma_{YY}(\bs_i, \bs_k; t)$ similarly.

Deriving the remainder of the forward equations for the moment-closure approximation to the spatial SIR jump process may be done in a similar fashion as shown in \eqref{eq:dmuxst_deriv}. In general, this is straightforward though tedious algebra. The full set of forward equations is provided in Appendix \ref{app:fe_complete}, with derivation details provided in Appendix \ref{app:fe_heuristic}. The alternative approach to deriving the forward equations is based on the original approach in \cite{whittle1957} and involves moment and cumulant generating functions. Details on that derivation method are provided in \ref{app:fe_whittle}. The two derivation methods yield the same forward equations.

We assume that the starting conditions of the outbreak do not correspond to a high probability of the infection dying out. This is a known source of moment-closure approximations failing and is sometimes referred to as ``epidemic fadeout'' \citep{lloyd2004}. We found in practice that checking the curves for the mean susceptibles to ensure they are monotonically nonincreasing served as a simple check on these assumptions. When the approximation holds, we are able to approximate well the population moments of the spatial SIR jump process. In Figure \ref{fig:examplemoments} we simulate 5,000 draws from the spatial SIR jump process for starting conditions with small epidemic-fadeout probability and compare them with the approximated moments from the moment-closure forward equations. We are able to approximate well the other moments also.

\section{Emulator for the mean and covariance functions}
\label{sec:lowrank}

The moment-closure approximation for the number of susceptibles $X(\bs,t)$ is a Gaussian process with mean function $\mu_X(\bs,t;\btheta)$ and spatial covariance function $\Sigma_{XX}(\bs_i,\bs_j;t,\btheta)$. Both of these functions vary over space and time and depend on the model parameters $\btheta = \left(\bbeta, \phi, S_0\right)^T$, where $\bbeta = \left(\beta(\bs_1),...,\beta(\bs_{n_s})\right)^T$ and $S_0 \in \mathcal{D}$ is the source of the outbreak. We assume that the recovery rate $\eta$ and starting time of the outbreak $T_0$ are known. Typically $\eta$ may be estimated using patient reports, and $T_0$ may be imputed based on initial reported cases. In this section we develop a statistical emulator to approximate the forward equations with an emulator that can be used for repeated function calls in an MCMC algorithm. To do so, we evaluate the forward equations of $K$ space-filling input parameters $\btheta_1,...,\btheta_K$ and then use the realizations of the forward equations to build a statistical prediction of their values at other inputs $\btheta^*$. We denote the output of running the forward equations for input $\btheta_k$ as $\bmu_X(\btheta_k) \in \mathbb{R}^{n_s \times n_t}$ and $\mathbf{\Sigma}_{XX}(\btheta_k) \in \mathbb{R}^{n_s \times n_s \times n_t}$, where $k \in \{1,...,K\}$. 

Much of the discussion in this section will use tensor terminology and notation. We provide a brief overview of the necessary background in Appendix \ref{app:tensors}. We first describe our methodology in Sections \ref{sec:lowrank_mean} and \ref{sec:lowrank_cov} in terms of the higher-order singular value decompositions (HOSVD) of tensors, as in \cite{gopalanwikle}. In Section \ref{sec:kriging} we discuss our imputation methodology for arbitrary $\btheta^* \notin \{\btheta_1,...,\btheta_K\}$. In Section \ref{sec:model_desc} we describe our model for new infections.

\subsection{Low-rank approximation to the mean function}
\label{sec:lowrank_mean}

We store our simulated output for the mean susceptibles in a third-order tensor $\mathcal{U} \in \mathbb{R}^{n_s \times n_t \times K}$, where the slice $\mathcal{U}_{.,.,k}$ is the matrix $\bmu_X(\btheta_k) \in \mathbb{R}^{n_s \times n_t}$. We construct a low-dimensional approximation to $\mathcal{U}$ using an HOSVD \citep{hosvdpaper, koldabader}, such that the low-dimensional approximation is $\hat{\mathcal{U}} \in \mathbb{R}^{J_s \times J_t \times K}$ with $J_s \leq n_s$ and $J_t \leq n_t$. The HOSVD is frequently compared to the well-known SVD for matrices, and it is a method to calculate a ``low-rank approximation [to a tensor] with small reconstruction error'' \citep{zare2018}.

We denote the spatial factor matrix of $\mathcal{U}$ as $\bgamma \in \mathbb{R}^{n_s \times J_s}$ and the temporal factor matrix of $\mathcal{U}$ as $\bdelta \in \mathbb{R}^{n_t \times J_t}$ with columns $\bgamma = \left[\bgamma_1, ..., \bgamma_{J_s}\right]$ and $\bdelta = \left[\bdelta_1, ..., \bdelta_{J_t}\right]$. This corresponds to a $J_s \times J_t \times K$ low-rank approximation to $\mathcal{U}$, with $J_s \leq n_s$ and $J_t \leq n_t$. The low-rank approximation to the mean susceptibles is therefore
\begin{equation}\label{eq:muxapprox}
   \mu_X(s,t;\btheta) \approx \sum_{i=1}^{J_s}\sum_{j=1}^{J_t}\gamma_i(s)\delta_j(t)m_{ij}(\btheta).
\end{equation}
The basis weights $m_{ij}(\btheta)$ control the dependence of the mean function on the model parameters $\btheta$ as well as the interactions between $\bgamma$ and $\bdelta$. These weights are the focus of our interpolation in Section \ref{sec:kriging}, and they result from calculating $\mathcal{U} \times_1 \bgamma^T \times_2 \bdelta^T$, where $\times_1$ and $\times_2$ are the n-mode products. We note that if $J_s = n_s$ and $J_t = n_t$, then our approximation to $\mathcal{U}$ is exact \citep{koldabader}. Finally, we implement streaming algorithms for this low-rank approximation for large $n_s$, $n_t$, and $K$ to avoid storing the entire tensor $\mathcal{U}$. See Appendix \ref{app:algorithms}.

\subsection{Low-rank approximation to the covariance function}
\label{sec:lowrank_cov}

Our approach to emulating the spatial covariance function is similar to our approach for the mean function. We store our simulated output for the spatial covariance matrices in a tensor $\mathcal{S} \in \mathbb{R}^{n_s \times n_s \times n_t \times K}$. The marginal spatial covariance matrix for $\btheta_k$ at time $t$ is therefore $\mathcal{S}_{.,.,t,k} = \mathbf{\Sigma}_{XX}(t,\btheta_k)$. We now wish to calculate a low-rank approximation $\hat{\mathcal{S}} \in \mathbb{R}^{L_s \times L_s \times L_t \times K}$ to $\mathcal{S}$ based on an HOSVD, where $L_s \leq n_s$ and $L_t \leq n_t$.

There are noteworthy differences in constructing an emulator for the covariance function for the susceptibles compared with the simpler emulator for the mean susceptibles. There is symmetry in $\mathcal{S}$ because the slice corresponding to $\mathcal{S}_{.,.,t,k}$ is a covariance matrix. Therefore, the factor matrices for the first and second mode are identical because the unfolded tensor is identical along its first and second modes. To ensure symmetry while avoiding redundancies, we build an emulator for $\mathbf{\Phi}(t,\btheta)$ such that $\mathbf{\Phi}(t,\btheta)\mathbf{\Phi}(t,\btheta)^T = \hat{\mathbf{\Sigma}}_{XX}(t,\btheta)$. It is more efficient to emulate $\mathbf{\Phi}(t,\btheta)$ because no matrix decompositions are required while model fitting.

We first calculate the spatial factor matrix $\bGamma = \left[\bGamma_1, ..., \bGamma_{L_s}\right] \in \mathbb{R}^{n_s \times L_s}$, where $\bGamma_1,...,\bGamma_{L_s}$ are the orthonormal spatial basis functions that capture spatial variability in the unfolded tensor along the first mode (equivalently, the second mode). We then calculate $\mathcal{Z} = \mathcal{S} \times_1 \bGamma^T \times_2 \bGamma^T$, which calculates time- and parameter-indexed weights for the interaction of the spatial basis functions. In the case of $L_s = n_s$, then we may write $\mathbf{\Sigma}_{XX}(t,\btheta) = \bGamma \mathcal{Z}_{.,.,t,k} \bGamma^T$. It follows there exists a Cholesky decomposition $\mathbf{C}(t,\btheta)$ such that $\mathbf{C}(t,\btheta) \mathbf{C}(t,\btheta)^T = \mathcal{Z}_{.,.,t,k}$, which naturally implies that we may set $\mathbf{\Phi}(t,\btheta) = \bGamma \mathbf{C}(t,\btheta)$ (see Appendix \ref{app:cholesky_pf}).

Approximating the Cholesky decompositions in an analogous fashion to an HOSVD worked well as a further dimension reduction. We found that doing a low-rank approximation to $\mathbf{C}(t,\btheta)$ works nearly as well as using $\mathbf{C}(t,\btheta)$ in terms of reconstructing $\mathbf{\Sigma}_{XX}(t,\btheta)$, but the computational run time is materially faster. We perform this low-rank approximation by forming a tensor $\mathcal{C} \in \mathbb{R}^{L_s \times L_s \times n_t \times K}$, where $\mathcal{C}_{.,.,t,k} = \mathbf{C}(t,\btheta)$, and calculating a temporal factor matrix for $\mathcal{C}$, i.e., we calculate the first $L_t$ left singular vectors of $\mathcal{C}$ unfolded along its third, temporal mode. We denote this temporal factor matrix as $\bDelta = \left[\bDelta_1,...,\bDelta_{L_t}\right]$, where $\bDelta_1,...,\bDelta_{L_t}$ are orthonormal temporal basis functions. We then calculate the weight tensor $\mathcal{M} \in \mathbb{R}^{L_s \times L_s \times L_t \times K}$ by calculating $\mathcal{C} \times_3 \bDelta^T$.

Therefore, we approximate the covariance function as
\begin{align}
\Phi_j(\bs, t; \btheta) =& \sum_{i = 1}^{L_s}\sum_{l = 1}^{L_t} \Gamma_i(\bs) \Delta_l(t) M_{ijl}(\btheta). \label{eq:sxx_tensor_approx_p2} \\
\mathbf{\Sigma}_{XX}(\bs, \bs'; t, \btheta) \approx& \sum_{j = 1}^{L_s} \Phi_j(\bs, t; \btheta) \Phi_j(\bs', t; \btheta) \label{eq:sxx_tensor_approx_p1}
\end{align}
We note that by modeling the covariance function as $\mathbf{\Phi}(t,\btheta) \mathbf{\Phi}(t,\btheta)^T$ with $\mathbf{\Phi}(t,\btheta) \in \mathbb{R}^{n_s \times L_s}$ comprised of elements $\Phi_j(s, t, \btheta)$, we ensure that our approximation for $\mathbf{\Sigma}_{XX}(t,\btheta)$ is non-negative definite for all $t$ and $\btheta$. We note as before that all our approximations are exact when $L_s = n_s$ and $L_t = n_t$ \citep{koldabader}, and we also note these calculations may be performed using streaming algorithms to avoid memory problems (see Appendix \ref{app:algorithms}).


\subsection{Gaussian process regression for interpolation}
\label{sec:kriging}

To fit our data model using MCMC, we need to repeatedly evaluate $\hat{\bmu}_X(\btheta^*)$ and $\hat{\mathbf{\Sigma}}_{XX}(\btheta^*)$ for proposed parameters $\btheta^*$. Given the low-rank approximations above, evaluating $\hat{\bmu}_X(\btheta^*)$ and $\hat{\mathbf{\Sigma}}_{XX}(\btheta^*)$ reduces to evaluating $m_{ij}(\btheta^*)$ and $M_{ijl}(\btheta^*)$. Therefore, we require a regression model for the weights as functions of the input parameters $\btheta$. 

To do so, we build a Gaussian-process regression for $m_{ij}(\btheta^*)$ and $M_{ijl}(\btheta^*)$ assuming independence between the indices $i, j, l$ and using the same covariance for all predictions. Here we describe our method to predict $m_{ij}(\btheta^*)$ given $\{m_{ij}(\btheta_1),...,m_{ij}(\btheta_K)\}$. We model $\text{Cov}\{m_{ij}(\btheta_{k_1}), m_{ij}(\btheta_{k_2})\} = \text{Mat\'ern}(d(\btheta_{k_1}, \btheta_{k_2}), \zeta, \kappa)$, where $d(\btheta_{k_1}, \btheta_{k_2}) = ||\btheta_{k_1} - \btheta_{k_2}||_2$, $\zeta$ is a range parameter, $\kappa$ is the smoothness parameter, and the Mat\'ern correlation function is defined as in \cite{cressiewikle} Section 4.1.1. We use a smoothness of $\kappa = 2.5$ for all our Mat\'ern correlation functions because the weights $m_{ij}$ and $M_{ijl}$ are smooth in the parameter space. We examine variograms to pick the range $\zeta$. Our prediction for $m_{ij}(\btheta^*)$ is then:
\begin{align}
    \hat{m}_{ij}(\btheta^*) &= \hat{\mu}_{ij} + \mathbf{E}_{21}\mathbf{E}_{11}^{-1}(\pmb{m^*} - \pmb{1} \hat{\mu}_{ij})
\end{align}
where $\pmb{m^*}$ is the vector of $m_{ij}(\btheta_k)$ for the $n_u$ nearest neighbors to $\btheta^*$, $\hat{\mu}_{ij}$ is the weighted mean of $\pmb{m^*}$, $\pmb{1}$ is a vector of ones $\in \mathbb{R}^{n_u}$, and the correlation matrices are $\mathbf{E}_{21} = \text{Cov}(m_{ij}(\btheta^*), \pmb{m^*})$ and $\mathbf{E}_{11} = \text{Cov}(\pmb{m^*})$ defined by the Mat\'ern correlation function. For further details, see \cite{cressieTextbook} Section 3.2. The full details of our approach are provided in Algorithms \ref{alg:mux_mcmc} and \ref{alg:sxx_mcmc} in Appendix \ref{app:algorithms}.


\subsection{Emulator model for new infections}
\label{sec:model_desc}

Using the emulators described above, our model for the new infections $y(\bs, t)$ is
\begin{subequations}
\begin{align}
    y(\bs,t) | \pmb{X}, p, \nu &\sim NB\left(p \left[X(\bs,t-1) - X(\bs,t)\right], \frac{p \left[X(s,t-1) - X(s,t)\right]}{\nu - 1}\right) \\
    \pmb{X}_t | \btheta, \bmu(t, \theta), \mathbf{\Phi}(t,\btheta), \balpha_t  &= \bmu_X(t, \btheta) + \mathbf{\Phi}(t, \btheta) \mathbf{B} \, \balpha_t \\
    \mu_X(\bs,t;\btheta) &= \sum_{i=1}^{J_s}\sum_{j=1}^{J_t}\gamma_i(s)\delta_j(t)\hat{m}_{ij}(\btheta) \\
    \Phi_j(\bs, t; \btheta) &= \sum_{i = 1}^{L_s}\sum_{l = 1}^{L_t} \Gamma_i(s) \Delta_l(t) \hat{M}_{ijl}(\btheta)
\end{align}
\end{subequations}
Under this parameterization, the mean of $y(\bs,t)$ is $p \left(X(\bs,t-1) - X(s,t)\right)$ with variance \\ $\nu\,p \left(X(\bs,t-1) - X(\bs,t)\right)$, with $p \in \left(0,1\right]$ being the reporting rate and $\nu > 1$ controlling overdispersion. These parameters model real-world variance in the number of new infections reported.

The forward equations do not model temporal correlation in the susceptibles. We use B-spline basis curves as a low-rank approximation to a time series to capture this temporal correlation. Let $\balpha_t = \{\alpha_{1,t},...,\alpha_{L_s,t}\}$, and let $\mathbf{B} \in \mathbb{R}^{n_t \times b}$ be a matrix containing $b$ B-spline basis curves standardized so that the diagonal entries of $\mathbf{B}\mathbf{B}^T$ are equal to 1. Under our model above then, $\mathbb{E}\left(\pmb{X}_t\right) = \bmu_X(t,\btheta)$ and $\mathbb{V}\left(\pmb{X}_t\right) = \mathbf{\Sigma}_{XX}(t,\btheta)$. 

We use a truncated normal prior for $\nu$ with mean 3, variance 25, and lower bound 1; uninformative normal priors for $\bbeta$ and $\phi$; a discrete uniform prior for $S_0$ over $\{1,...,n_s\}$; and a standard normal prior for $\balpha$. We provide suggestions on implementation details in Appendix \ref{app:implementation}.


\section{Simulation study}
\label{sec:sim_study_all}

We analyze the performance of our method on data simulated from the spatial SIR jump process described in Section \ref{sec:spatialsir_jump}. We simulate counts of daily susceptibles from this jump process using the Gillespie stochastic simulation algorithm \citep{gillespie1976, gillespie1977}. The simulated response data $y(\bs,t)$ are generated by drawing from an overdispersed negative-binomial distribution with mean ${p \{X(s,t-1) - X(s,t)\}}$ and with overdispersion $\nu > 1$.

\subsection{Simulation study with constant $\beta$}
\label{sec:sim_study_1}

We define a square grid of 25 spatial locations with population size of 100,000, and we set the center of the square grid as $S_0$. The neighborhood structure is defined as in a regular lattice with rook adjacencies. We assume 100 individuals were infected at time 0 at $S_0$ and that the disease had not spread yet to any other locations. We use $\beta = 0.043 \,\,\forall s \in \mathcal{D}$. We also use $\phi = 0.025$, $\eta = 0.019$, $n_t = 80$, and $\nu = 3.2$. We begin our analysis at day 61, so $t \in \{61,...,140\}$, reflecting data not typically being collected in the early days of an outbreak. We simulate 200 data sets for each simulation study.

We tune our emulators to use $J_s = 20$, $J_t = 10$, $L_s = 10$, and $L_t = 10$ (see Appendix \ref{app:sim1} for details). For the first simulation, we use $p = 1$ so there is no underreporting, and we use $\nu = 3.2$. For the second simulation, we use $p = 0.75$ so that not all new infections get reported. For the third simulation, we use $p = 1$ but set $\nu = 10$ to more than triple overdispersion. In all three cases, we also consider fitting a simpler model that only fits the mean curves, i.e., sets all $\balpha$ equal to 0. Table \ref{tab:sim_study_results_1} shows our results.

\begin{table}[H]
\caption{Results for first set of simulation studies using the full emulator approximation to the moment-closure equations or using only the mean emulator (``mean only''). Empirical credible-interval coverage is shown along with MSE and its standard error, multiplied by $10^6$.} \label{tab:sim_study_results_1}
\begin{center}
\begin{tabular}{ cc|cc|cc|cc }
Scenario & Model & \makecell{$\beta$, 95\% \\ cov.} & \makecell{$\beta$, 99\% \\ cov.} & \makecell{$\phi$, 95\% \\ cov.} & \makecell{$\phi$, 99\% \\ cov.} & \makecell{$\beta$, MSE \\ (SE)} & \makecell{$\phi$, MSE \\ (SE)} \\
\hline
Baseline & Full & 96.0\% & 99.0\% & 86.0\% & 96.5\% & \makecell{3.507 \\ (0.366)} & \makecell{0.547 \\ (0.058)}  \\
Baseline & Mean only & 41.0\% & 51.5\% & 31.0\% & 38.5\% & \makecell{26.153 \\ (5.375)} & \makecell{3.363 \\ (0.590)} \\
\hline
$p = 0.75$ & Full & 96.0\% & 99.0\% & 89.0\% & 95.5\% & \makecell{4.523 \\ (0.434)} & \makecell{0.595 \\ (0.058)} \\
$p = 0.75$ & Mean only & 46.5\% & 60.0\% & 38.0\% & 48.5\% & \makecell{15.005 \\ (1.644)} & \makecell{2.018 \\ (0.217)} \\
\hline
$\nu = 10$ & Full & 95.5\% & 97.0\% & 92.0\% & 96.5\% & \makecell{7.873 \\ (0.819)} & \makecell{0.832 \\ (0.081)} \\
$\nu = 10$ & Mean only & 63.0\% & 74.0\% & 49.5\% & 61.0\% & \makecell{18.574 \\ (2.934)} & \makecell{2.354 \\ (0.321)} \\
\end{tabular}
\end{center}
\end{table}

As Table \ref{tab:sim_study_results_1} shows, a key component to our good coverage and parameter estimates is the inclusion of the covariance emulator. Fitting only the mean curves results in significantly worse coverage. These worse results are caused by failing to allow sufficient model flexibility for more infections or fewer infections than expected. However, when the covariance emulator is included, we are able to estimate the parameters well with good coverage.

\subsection{Simulation study with spatially varying $\beta(s)$}
\label{sec:sim_study_2}

For our second set of simulations, we allow $\beta$ to vary spatially such that $\beta(\bs) = \exp\{\beta_0 + \beta_1 x(\bs)\}$. We draw $\exp\{x^*(\bs)\} \overset{iid}{\sim} U(0, 3000) \,\forall \bs \in \mathcal{D}$ and then calculate $x(\bs) = x^*(\bs) - \bar{x^*}(\bs)$ to center the spatially-varying covariate. We use $\beta_0 = -2.83$, $\beta_1 = 0.1$, $\phi = 0.045$, $\eta = 0.04$, $\nu = 2.5$, and $p = 1$ while keeping $n_s = 25$ and $n_t = 80$. We begin our analysis at day 21, i.e., $t \in \{21, ..., 100\}$. We again use $J_s = 20$, $J_t = 10$, $L_s = 10$, and $L_t = 10$ (see Appendix \ref{app:sim2}).

Table \ref{tab:sim_study_results_2_cov} shows our coverage results compared with the same model run with only the mean emulator, and Table \ref{tab:sim_study_results_2_rmse} similarly shows our MSEs for $\beta_0$, $\beta_1$, and $\phi$. With the addition of a spatial covariate, the difference between including or excluding the covariance emulator is striking. The ``only means'' model performed poorly, and many of the chains became stuck. The problem is there is not enough flexibility in the model to estimate the parameters well. However, with the inclusion of the covariance emulator, we again get good coverage and MSEs.

\begin{table}[H]
\caption{Empirical credible-interval coverage results for second set of simulation studies using the full emulator approximation to the moment-closure equations or using only the mean emulator (``mean only''). In this study $\beta(\bs) = \exp\{\beta_0 + \beta_1 x(\bs)\}$ where $x(\bs)$ is a spatially varying covariate.}
\label{tab:sim_study_results_2_cov}
\begin{center}
\begin{tabular}{ c|cc|cc|cc }
Study & \makecell{$\beta_0$, 95\% \\ cov.} & \makecell{$\beta_0$, 99\% \\ cov.} & \makecell{$\beta_1$, 95\% \\ cov.} & \makecell{$\beta_1$, 99\% \\ cov.} & \makecell{$\phi$, 95\% \\ cov.} & \makecell{$\phi$, 99\% \\ cov.} \\
\hline
Full model & 92.0\% & 96.0\% & 89.5\% & 96.0\% &  88.5\% & 95.5\% \\
Only means & 24.5\% & 27.5\% & 14.0\% & 17.0\% & 21.5\% & 28.0\% \\
\end{tabular}
\end{center}
\end{table}

\begin{table}[H]
\caption{MSEs and their standard errors for second set of simulation studies using the full emulator approximation to the moment-closure equations or using only the mean emulator (``mean only''). In this study $\beta(\bs) = \exp\{\beta_0 + \beta_1 x(\bs)\}$ where $x(\bs)$ is a spatially varying covariate. Both MSEs and SEs are multiplied by $10^4$.}
\label{tab:sim_study_results_2_rmse}
\begin{center}
\begin{tabular}{ c|c|c|c }
Study & \makecell{$\beta_0$, MSE \\ (SE)} & \makecell{$\beta_1$, MSE \\ (SE)} & \makecell{$\phi$, MSE \\ (SE)} \\
\hline
Full model & \makecell{25.202 \\ (2.843)} & \makecell{1.298 \\ (0.130)} & \makecell{0.010 \\ (0.001)} \\
\hline
Only means & \makecell{171.197 \\ (7.682)} & \makecell{7.682 \\ (1.301)} & \makecell{0.071 \\ (0.005)} \\
\end{tabular}
\end{center}
\end{table}


\section{Data analysis}
\label{sec:brazil}

We demonstrate the application of our model to Zika virus outbreak data in Brazil. The Zika virus is for most humans a relatively mild virus, causing fevers, rashes, and various pains \citep{cdczika1}. Unfortunately, there are serious complications for pregnant women, whose children may be born with multiple severe birth defects including microcephaly \citep{cdczika2}. The spread of a virus dangerous to fetuses coincided with the 2016 summer Olympics held in Rio de Janeiro, and official guidance typically recommended pregnant women not attend the Olympics that year \citep{cdczika3}.

\begin{figure}[H]
\caption{Weekly new outbreaks of the Zika in Brazil by state, late 2015 into first half of 2016. The new infections are normalized in the plot by state population for display purposes. Both the intensity and the timing of the outbreaks vary by location.}
\label{fig:em_brazil}
\begin{center}
\begin{tabular}{c}
\includegraphics[width=0.75\textwidth]{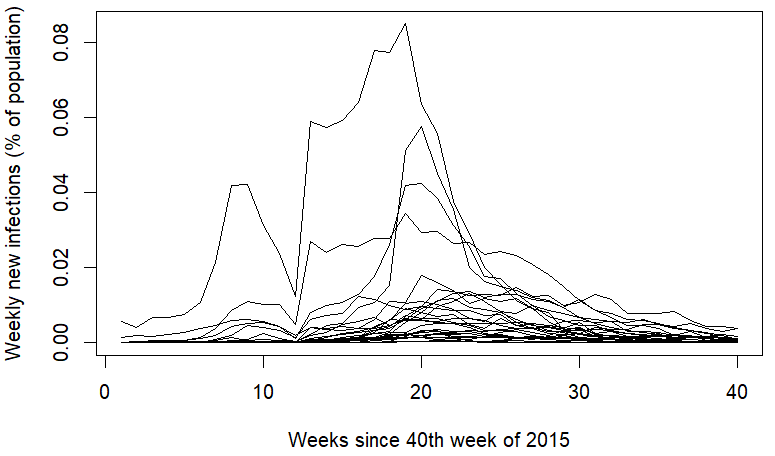}
\end{tabular}
\end{center}
\end{figure}

The Pan American Health Organization (PAHO) reports government-supplied data on weekly new cases of the Zika virus during this time period \citep{zikadata}. The data are reported for each of the 27 states of Brazil, including the Federal District. The first significant outbreak began in the second half of 2015 and continued into 2016, though the first cases in Brazil began in the state of Maranhao in early 2015 \citep{pahoZikaTimeline}. PAHO reports weekly cases during this period. Figure \ref{fig:em_brazil} plots the reported new cases beginning in the 41st week of 2015 continuing through the 28th week of 2016, corresponding roughly to October 2015 through July 2016 \citep{zikadata}. The counts have been normalized by the population sizes at each spatial location. 

\noindent We fit our model to these 40 weeks of outbreak data. 

A recent paper modeled the spread of Zika virus using a compartmental model that incorporated spread by mosquitoes \citep{sadeghieh2021}. We rely on their estimate of a weekly $\gamma$, which was 1.2. We also expected the reported weekly new infections to be underreported. One paper suggests that the reporting rate for Zika infections was somewhere in 7\% to 17\% \citep{shuttmanore2017}, though that was for Central and South America as a whole. It is reasonable to expect this varied materially by country. To pick starting values for our model and to evaluate reporting rates, we fit a simple model to the Zika data with a Poisson likelihood and mean function analogous to a deterministic model of our spatial SIR jump process. We assumed the outbreak began in the first week of 2015 in Maranhao. We found that even the lower bound of 7\% yielded poor fits to the data, and we estimated that the reporting rates were often closer to 1\% and varied by state. After estimating the reporting rates in the preliminary analysis, we fixed those reporting rates for all subsequent analysis. We model $\beta(\bs) = \exp\{\beta_0 + \beta_1 x(\bs)\}$, where $x(\bs)$ is the scaled and centered log population density of state $\bs$ \citep{IBGE}.

Using the starting values from our preliminary analysis, we prepared our emulator simulations. We found we needed both a wide coverage of the parameter space to account for parameter uncertainty, but we also needed a relatively fine mesh on the parameter space. We used $K = 100,000$ over a wide parameter space, specifically $\beta_0 \in \left(-0.015, 0.000\right)$, $\beta_1 \in \left(0.13, 0.28\right)$, and $\phi \in \left(0.055, 0.145\right)$. Those ranges and values were first based on the preliminary analysis and then based on running short, small-$K$ emulator-based models to make adjustments as necessary (in particular, to prevent the MCMC chains from hitting emulator-space boundaries, as discussed in Web Appendix D). We discarded $20,727$ of the proposed parameter values because they corresponded to implausible combinations of $\left(\beta_0, \beta_1, \phi\right)^T$ based on fadeouts happening too quickly and the moment-closure approximation not holding, as determined by examining the derivatives of the mean susceptible curves. These all corresponded to cases where $\beta_1$ was high and $\phi$ was low, which represent scenarios with extreme outbreaks in urban states that do not spread to neighboring states. We used $J_s = 20$, $J_t = 10$, $L_s = 10$, and $L_t = 10$, but we increased the count of nearest neighbors used for the imputations by kriging to 20 for both the mean and covariance emulator. This was to help with the large parameter space relative to $K$. As stated before, we assume $S_0$ was Maranhao, $T_0$ was the first week of January, and $Y(S_0, T_0) = 100$.

We estimate $\hat{\beta}_0 = -0.077 \left(-0.109, -0.052\right)$, $\hat{\beta}_1 = 0.202 \left(0.193, 0.213\right)$, and $\hat{\phi} = 0.087 \left(0.082, 0.092\right)$, where the estimates are posterior means and the numbers in parentheses are the 95\% credible intervals. 

There are two implications from our model fit. First is that, as expected, higher population densities are associated with higher rates of transmission. The second is that there is evidence of strong spatial spread of Zika as evidenced by the credible interval for $\phi$. We note that the parameter estimates and credible intervals were not in the emulator design space corresponding to problematically low $\phi$ and high $\beta_1$ as discussed previously.

\begin{figure}[H]
\caption{Plots of reported new Zika infections in late 2015 through early 2016 in Rio de Janeiro (L) and Bahia (R), the two states with the most reported new infections. The lines represent the fitted mean curves.}
\label{fig:brazilfits_good}
\begin{center}
\begin{tabular}{cc}
\includegraphics[width=0.46\textwidth]{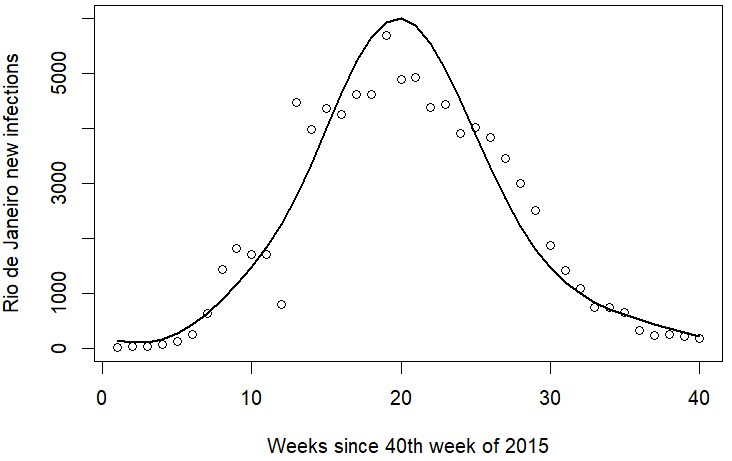} &
\includegraphics[width=0.46\textwidth]{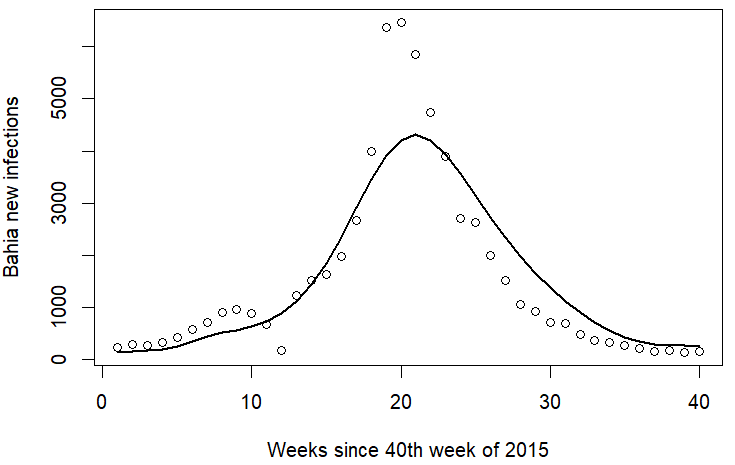}
\end{tabular}
\end{center}
\end{figure}

Our model fit different states better than others. In particular, our model fit best in states with the highest counts of outbreaks. In Figure \ref{fig:brazilfits_good} we plot the new counts of Zika infections in Rio de Janeiro and Bahia, the two states with the highest counts of reported new infections.

\begin{figure}[H]
\caption{States with smaller numbers of new infections have poorer fits, such as Piaui (L). Examining model discrepancy using 1 - L2 norms of the observations and the fit mean curves suggests the center and northeast regions have poorer fits (R, norms standardized by L2 norms of state infections).}
\label{fig:brazilfits_poor}
\begin{center}
\begin{tabular}{c}
\includegraphics[width=0.95\textwidth]{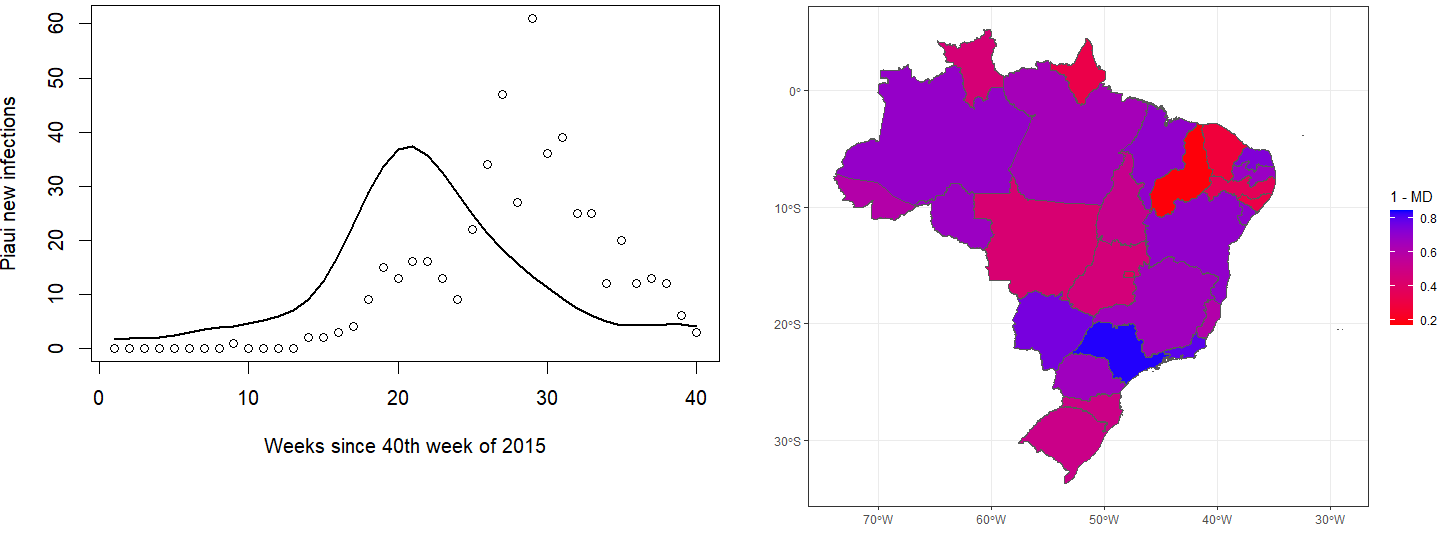}
\end{tabular}
\end{center}
\end{figure}

Our model fits worse in areas with smaller number of infections, and it fits worse in the northeast states. These sometimes coincide, such as in Piaui where only 497 infections were reported during this period out of a population of 3.2 million and where our model fits more poorly (Figure \ref{fig:brazilfits_poor}, left). To look for spatial patterns in model discrepancy, we calculated $1 - ||\by_s-\blambda_s||_2 / ||\by_s||_2$, where $\by_s$ is the time series of new infections at site $\bs$ and $\blambda_s$ is the time series of the mean curves at site $\bs$ (Figure \ref{fig:brazilfits_poor}, right). This graph suggests that the northeast region in particular has a poorer fit, though it is noticeably better in Rio Grande do Norte and Paraiba (the two purple states in the northeast of Brazil). The fit in the center region, in particular Mato Grosso, also is not as good as the more urban regions on the eastern coast.

\noindent We considered incorporating a model-discrepancy term into our model. Our preliminary results, which we describe in Web Appendix H, suggest we would get similar parameter and uncertainty estimates for $\beta_0$, $\beta_1$, and $\phi$.

After conducting our analysis, we increased the number of spatial basis functions for both the mean and covariance emulators to assess the sensitivity of our model. Specifically, we considered increasing $J_s$ to 25 and separately $L_s$ to 15 to assess if more basis functions may better capture spatial variability in the outbreaks and to help assess how sensitive our results are to the low-rank approximation. Our results were insensitive to the increase in $J_s$ and $L_s$. Table \ref{tab:brazil_results_sens} in Appendix \ref{app:zika} shows 95\% credible intervals for increasing the number of spatial basis functions for the emulators. Results were similar for the point estimates as well as for the measures of model discrepancy, as described previously.

In conclusion, our results show there is clear evidence of spatial spread, and our preliminary analysis suggests the reporting rates for new Zika cases during this time period may have been much lower than expected. There is still some clear model misspecification, suggesting that more features may be needed. Alternatively, incorporating spread by mosquitoes from a model like in \cite{sadeghieh2021} along with our spatial spread may yield the best possible results.


\section{Discussion}
\label{sec:discussion}

We have proposed a novel method to analyze spatial infection data based on combining moment-closure approximations with emulators. Both components of our method are vital. The emulator approach is needed because of the computational run time of the forward equations. However, without the forward equations the emulator-based approximation would not work as well. In lieu of using a moment-closure approximation to the spatial SIR jump process, the alternative is to simulate many draws from the stochastic process to calculate sample means and spatial covariances. This approach is slower than running the forward equations, and the resulting sample moments would be subject to sampling variability. This means the weights $m_{ij}(\btheta)$ and $M_{ijl}(\btheta)$ would not be as smooth in the parameter space as they are with the deterministic forward equations.

Although our framework has been presented in light of the spatial SIR jump process, it can be extended to other continuous-time Markov processes. This can allow potentially easier statistical analysis in domains such as in the veterinary literature, which often relies on repeated simulations of complex compartmental models for epidemiological studies \citep[e.g.,][]{galvis2022cshl, jones2021}. Our method could allow a significant reduction in computational time to fit these models.

There are several areas where there is room improvement in our approach. The first is that we assume that there is no epidemic fadeout. Research into more concrete conditions for spatial moment-closure approximations to work well will be in important to generalize our approach. The second is we assume $\eta$, $p$, and the starting time of the infection are known or could be estimated in preliminary analysis, all of which may be able to be relaxed. The third is practical: our approach requires a significant amount of hands-on work, requiring the user to derive forward equations, tune emulator designs, and then tune an MCMC algorithm. Finding ways to simplify the hands-on work would allow for wide adoption of our methodology.

\bibliographystyle{apalike}
\bibliography{jpt-refs}

\appendix


\section{Background on SIR methodologies for a closed population}
\label{app:bground}

We review the basics of non-spatial SIR models. We start with the deterministic SIR model before then reviewing the stochastic SIR model, including introducing the moment-closure approximation to the stochastic SIR model proposed in \cite{isham1991}.

\subsection{Deterministic SIR model}
\label{app:bground_deterministic}

The SIR model is a compartmental model that describes the transmission of a disease through a single, closed population \citep{kermackmckendrick, tollesluong}. The population is divided into three groups, or compartments: the ``susceptibles'' are the members of the population who have not been infected, the ``infectious'' are those who have contracted the disease and are capable of spreading it to others, and the ``recovered''  are those who were once but are no longer infectious (including those who have died from the disease). There are numerous extensions and variations on this basic premise, including separating the ``exposed'' members (those who have contracted a disease but are not yet able to spread it to others) from the infectious members \citep[see, e.g.,][]{seirexample1, seirexample2}.

The SIR model is typically described in terms of the differential equations that govern the changes in the compartments through time \citep{allen2008}. Let $X(t)$ refer to the number of susceptibles at time $t$, $Y(t)$ refer to the number of infectious at time $t$, and $Z(t)$ refer to the number of recovered at time $t$. The population size $N$ is constant and is equal to $X(t) + Y(t) + Z(t)$ for any $t$. The differential equations for the SIR model are:
\begin{align}
\begin{split}
    \frac{dX(t)}{dt} &= - \frac{\beta X(t) Y(t)}{N} \\
    \frac{dY(t)}{dt} &= \frac{\beta X(t) Y(t)}{N} - \eta Y(t) \\
    \frac{dZ(t)}{dt} &= \eta Y(t) \label{eq:sir_diffeq}
\end{split}
\end{align}
where $\beta$ is a parameter measuring the infectiousness of the disease and $\eta$ is a parameter controlling the rate at which infectious individuals recover.

Given $\beta$ and $\eta$ and initial conditions $X(0), Y(0),$ and $Z(0)$, the differential equations yield smooth, continuous curves. Relying on these curves may be reasonable for modeling the mean number of susceptibles, infectious, and recovered, but these deterministic equations may not reflect reality well for reasons similar to those discussed in Section 1.1.2. We instead focus herein on stochastic SIR models, also known as the general stochastic epidemic model in the context of a single, closed population \citep{allen2008, isham1991}.

\subsection{Stochastic SIR model and its moment-closure approximation}
\label{app:bground_stochastic}

Consider a small increment in time $\Delta t$ that is small enough that only three events can occur: a susceptible person becomes infectious, an infectious person recovers, or neither happens \citep{allen2008}. Suppressing an explicit description of $Z(t)$, the transition probabilities are:
\begin{align}
\begin{split}
    P(X(t + \Delta t) = X(t) - 1, Y(t + \Delta t) = Y(t) + 1 | X(t), Y(t)) &\approx \beta X(t)Y(t) \Delta t / N \\
    P(X(t + \Delta t) = X(t), Y(t + \Delta t) = Y(t) - 1 | X(t), Y(t)) &\approx \eta Y(t) \Delta t \\
    P(X(t + \Delta t) = X(t), Y(t + \Delta t) = Y(t) | X(t), Y(t)) &\approx 1 - (\beta X(t) Y(t) / N \\& + \eta Y(t))\Delta t \label{eq:sir_jump_probs}
\end{split}
\end{align}
See \cite{allen2008} for more details. We refer to this stochastic model and similar stochastic models as ``SIR jump processes'' in this and the next chapter.

Though the stochastic model is more realistic than the simpler deterministic model, it can be analytically intractable to use for parameter inference with real data. This challenge has led to research into approximating the nonlinear SIR jump process with a simpler process such as a Gaussian process \citep[for a review of the literature, see][]{buckinghamjeffery2018}. We adopt the approach of \cite{isham1991} to approximate the SIR jump process with a Gaussian process based on a moment-closure approximation. The justification for this approximation may be argued by considering the Gaussian process as a limiting distribution as the population size increases \citep{kurtz1970, kurtz1971, lloyd2004}. Here, this approximation means modeling $X(t)$ and $Y(t)$ as being jointly normal at time $t$ and then modeling how their means and covariances evolve through time. Let $\mu_X(t)$ be the mean number of susceptibles at time $t$, $\mu_Y(t)$ be the mean number of infectious at time $t$, $\sigma_{XX}(t)$ and $\sigma_{YY}(t)$ be the marginal variances for the susceptibles and infectious (respectively) at time $t$, and $\sigma_{XY}(t)$ be the covariance between the susceptibles and infectious at time $t$. The five differential equations are:
\begin{align}
\begin{split}
    \frac{d \mu_X(t)}{dt} =& - \frac{\beta}{N} \bigl(\mu_X(t) \mu_Y(t) + \sigma_{XY}(t)\bigr)  \\
    \frac{d \mu_Y(t)}{dt} =& \frac{\beta}{N} \bigl(\mu_X(t) \mu_Y(t) + \sigma_{XY}(t)\bigr) - \eta \mu_Y(t) \\
    \frac{d \sigma_{XX}(t)}{dt} =& \frac{-2 \beta}{N} \bigl(\mu_X(t) \sigma_{XY}(t) + \mu_Y(t) \sigma_{XX}(t)\bigr) + \frac{\beta}{N} \bigl(\mu_X(t) \mu_Y(t) + \sigma_{XY}(t)\bigr) \\
    \frac{d \sigma_{XY}(t)}{dt} =& \frac{\beta}{N} \bigl(\mu_X(t) \sigma_{XY}(t) - \mu_X(t) \sigma_{YY}(t) + \mu_Y(t) \sigma_{XX}(t) - \mu_Y(t) \sigma_{XY}(t) \\
    &- \mu_X(t) \mu_Y(t) - \sigma_{XY}(t)\bigr) - \eta \sigma_{XY}(t) \\
    \frac{d \sigma_{YY}(t)}{dt} =& \frac{2 \beta}{N} \bigl(\mu_X(t) \sigma_{YY}(t) + \mu_Y(t) \sigma_{XY}(t)\bigr) + \frac{\beta}{N} \bigl(\mu_X(t) \mu_Y(t) + \sigma_{XY}(t)\bigr) \\
    &- 2\eta \sigma_{YY}(t) + \eta \mu_Y(t) \label{eq:ishamforward}
\end{split}
\end{align}
We refer to differential equations for the moments of a Gaussian process such as in \eqref{eq:ishamforward} above as forward equations.

There are two ways of deriving equations \eqref{eq:ishamforward}. The first way follows the approach outlined in \cite{whittle1957} and relies on manipulating and deriving the derivatives of moment-generating functions of the stochastic process. Alternatively, these forward equations can also be derived by following a heuristic approach using the probabilities defined in \eqref{eq:sir_jump_probs}. Consider the first equation in \eqref{eq:ishamforward}, the change in the mean number of susceptibles with respect to time. For a sufficiently small but otherwise arbitrary time step $\Delta t$, we would expect $X(t + \Delta t) = X(t) - 1$ to happen with probability $\beta X(t) Y(t) \Delta t / N$. Taking the expectation of $[X(t)-1-X(t)]\beta X(t) Y(t) \Delta t / N$ with respect to $X(t)$ and $Y(t)$ being jointly bivariate normal and then taking the limit as $\Delta t \rightarrow 0$ yields the equation for $\frac{d\mu_X(t)}{dt}$ in \eqref{eq:ishamforward}.

\section{Details on spatial SIR jump process forward equations}
\label{app:fe}

We provide details on the moment-closure approximation to the spatial SIR jump process. First, we provide the full set of forward equations. Next, we provide additional details and examples on deriving the forward equations using the heuristic approach. Finally, we show how to use the original approach taken in \cite{whittle1957} to derive the forward equations, beginning with an example of re-deriving the forward equations from \cite{isham1991}.

\subsection{Complete set of forward equations for spatial SIR jump process approximation}
\label{app:fe_complete}

The forward equations are:
\begin{align}
    \frac{d\mu_X(\bs_i,t)}{dt} =& -\frac{\beta(\bs_i)}{N(\bs_i)}\bigl(\mu_X(\bs_i,t) \mu_Y(\bs_i,t) + \sigma_{XY}(\bs_i, \bs_i; t)\bigr) \notag\\
    &- \frac{\phi}{N(\bs_i)}\sum_{\bs_k\in\mathcal{N}(\bs_i)}\bigl(\mu_X(\bs_i,t)\mu_Y(\bs_k,t) + \sigma_{XY}(\bs_i,\bs_k; t)\bigr) \\\notag\\
\frac{d\mu_Y(\bs_i)}{dt} =& -\frac{d\mu_X(\bs_i)}{dt} - \eta \mu_Y(\bs_i) \\\notag\\
        \frac{d\sigma_{XX}(\bs_i,\bs_i)}{dt} =& \frac{d\mu_X(\bs_i)}{dt} - \frac{2 \beta(\bs_i)}{N(\bs_i)}\bigl(\mu_Y(\bs_i)\sigma_{XX}(\bs_i,\bs_i) + \mu_X(\bs_i)\sigma_{XY}(\bs_i,\bs_i)\bigr) \notag \\
    & - \frac{2\phi}{N(\bs_i)}\sum_{\bs_k\in \mathcal{N}(\bs_i)}\left[\mu_Y(\bs_k)\sigma_{XX}(\bs_i,\bs_i) + \mu_X(\bs_i)\sigma_{XY}(\bs_i,\bs_k)\right] \\\notag\\ 
\frac{d\sigma_{YY}(\bs_i,\bs_i)}{dt} =& -\frac{d\mu_X(\bs_i)}{dt} + \frac{2\beta(\bs_i)}{N(\bs_i)}\bigl(\mu_X(\bs_i)\sigma_{YY}(\bs_i,\bs_i) + \mu_Y(\bs_i)\sigma_{XY}(\bs_i,\bs_i)\bigr) \notag \\
    & + \frac{2\phi}{N(\bs_i)}\sum_{\bs_k \in \mathcal{N}(\bs_i)}\left[\mu_X(\bs_i)\sigma_{YY}(\bs_i,\bs_k)+\mu_Y(\bs_k)\sigma_{XY}(\bs_i,\bs_i)\right] \notag \\
    &+ \eta\bigl(\mu_Y(\bs_i) - 2\sigma_{YY}(\bs_i,\bs_i)\bigr) \\\notag\\
\frac{d\sigma_{XY}(\bs_i,\bs_i)}{dt} =& \frac{d\mu_X(\bs_i)}{dt} -\frac{1}{2}\left(\frac{d\sigma_{XX}(\bs_i,\bs_i)}{dt} + \frac{d\sigma_{YY}(\bs_i,\bs_i)}{dt}\right) \notag \\
    & + \eta \left(2\sigma_{YY}(\bs_i,\bs_i) - \mu_Y(\bs_i) - \sigma_{XY}(\bs_i,\bs_i)\right)
\end{align}
\newpage
\begin{align}
\frac{d\sigma_{XX}(\bs_i, \bs_j)}{dt} =& -\frac{\beta(\bs_i)}{N(\bs_i)}(\mu_X(\bs_i)\sigma_{XY}(\bs_j,\bs_i) + \mu_Y(\bs_i)\sigma_{XX}(\bs_i,\bs_j)) \notag \\
    & -\frac{\beta(\bs_j)}{N(\bs_j)}(\mu_X(\bs_j)\sigma_{XY}(\bs_i,\bs_j) + \mu_Y(\bs_j)\sigma_{XX}(\bs_i, \bs_j))\notag \\ 
    & - \frac{\phi}{N(\bs_i)}\sum_{\bs_k \in \mathcal{N}(\bs_i)}\left[\mu_X(\bs_i)\sigma_{XY}(\bs_j, \bs_k) + \mu_Y(\bs_k)\sigma_{XX}(\bs_i,\bs_j)\right]\notag \\
    & - \frac{\phi}{N(\bs_j)}\sum_{\bs_l \in \mathcal{N}(\bs_j)}\left[\mu_X(\bs_j)\sigma_{XY}(\bs_i, \bs_l) + \mu_Y(\bs_l)\sigma_{XX}(\bs_i,\bs_j)\right]\\\notag\\
\frac{d\sigma_{YY}(\bs_i, \bs_j)}{dt} =& \frac{\beta(\bs_i)}{N(\bs_i)}(\mu_X(\bs_i)\sigma_{YY}(\bs_i,\bs_j) + \mu_Y(\bs_i)\sigma_{XY}(\bs_i, \bs_j))\notag \\
    & + \frac{\beta(\bs_j)}{N(\bs_j)}(\mu_X(\bs_j)\sigma_{YY}(\bs_j, \bs_i) + \mu_Y(\bs_j)\sigma_{XY}(\bs_j, \bs_i))\notag \\
    & + \frac{\phi}{N(\bs_i)}\sum_{\bs_k \in \mathcal{N}(\bs_i)}\left[\mu_X(\bs_i)\sigma_{YY}(\bs_k, \bs_j) + \mu_Y(\bs_k)\sigma_{XY}(\bs_i, \bs_j)\right]\notag \\
    & + \frac{\phi}{N(\bs_j)}\sum_{\bs_l \in \mathcal{N})(\bs_j)}\left[\mu_X(\bs_j)\sigma_{YY}(\bs_l, \bs_i) + \mu_Y(\bs_l)\sigma_{XY}(\bs_j, \bs_i)\right]\notag \\
    & - 2 \eta \sigma_{YY}(\bs_i, \bs_j) \\\notag\\
    \frac{d\sigma_{XY}(\bs_i, \bs_j)}{dt} =& - \frac{\beta}{N(\bs_i)}(\mu_X(\bs_i) \sigma_{YY}(\bs_i, \bs_j) + \mu_Y(s_i)\sigma_{XY}(\bs_i, \bs_j))\notag \\
    & + \frac{\beta}{N(\bs_j)}(\mu_X(\bs_j) \sigma_{XY}(\bs_i, \bs_j) + \mu_Y(\bs_j)\sigma_{XX}(\bs_i, \bs_j))\notag \\
    & - \frac{\phi}{N(\bs_i)}\sum_{\bs_k \in \mathcal{N}(\bs_i)}\left[ \mu_X(\bs_i)\sigma_{YY}(\bs_j, \bs_k) + \mu_Y(\bs_k)\sigma_{XY}(\bs_i, \bs_j) \right]\notag \\
    & + \frac{\phi}{N(\bs_j)}\sum_{\bs_l \in \mathcal{N})(\bs_j)}\left[ \mu_X(\bs_j)\sigma_{XY}(\bs_i, \bs_l) + \mu_Y(\bs_l) \sigma_{XX}(\bs_j, \bs_i) \right]\notag \\
    & - \eta \sigma_{XY}(\bs_i, \bs_j)
\end{align}

\subsection{Details on heuristic approach to deriving forward equations}
\label{app:fe_heuristic}

We now provide additional math from Section 2.3.2 to derive the forward equations heuristically. We begin by proving two expectation identities that are used repeatedly throughout these derivations. Assuming $X$ and $Y$ are jointly normal with standard notation to indicate the means and variances, then:
\begin{align*}
    \mathbb{E}\{X^2 Y\} =& \mathbb{E}_Y\{\mathbb{E}_{X|Y}\{X^2 y | Y = y\}\} \\
    =& \mathbb{E}_Y\{Y \bigl(\sigma_{XX}-\frac{\sigma_{XY}^2}{\sigma_{YY}} + (\mu_X + \sigma_{XY}\sigma_{YY}(y-\mu_Y))^2\bigr)\} \\
    =& \mu_X \sigma_{XX} - \mu_Y \frac{\sigma_{XY}^2}{\sigma_{YY}} + \mu_X^2 \mu_Y + \frac{\sigma_{XY}^2}{\sigma_{YY}^2} \mathbb{E}_Y\{(Y-\mu_Y)^2 Y\} \\ &+ 2\mu_X \frac{\sigma_{XY}}{\sigma_{YY}}\mathbb{E}_Y\{Y (Y-\mu_Y)\} \\
    =& \mu_X \sigma_{XX} - \mu_Y \frac{\sigma_{XY}^2}{\sigma_{YY}} + \mu_X^2 \mu_Y + \frac{\sigma_{XY}^2}{\sigma_{YY}^2} \bigl(\mathbb{E}_Y(Y^3)+\mu_Y^3 - 2\mu_Y \mathbb{E}_Y\{Y^2\}\bigr) + 2\mu_X \sigma_{XY} \\
    =& \mu_X \sigma_{XX} - \mu_Y \frac{\sigma_{XY}^2}{\sigma_{YY}} + \mu_X^2 \mu_Y + \frac{\sigma_{XY}^2}{\sigma_{YY}^2} (\mu_Y \sigma_{YY}) + 2\mu_X \sigma_{XY} \\
    =& \mu_Y \sigma_{XX} + \mu_X^2 \mu_Y + 2 \mu_X \sigma_{XY}
\end{align*}
where the second equality comes from $\mathbb{V}(X) + \mathbb{E}(X)^2 = \mathbb{E}(X^2)$ and the properties of conditional normal distributions.

We also will frequently use $\mathbb{E}(XYZ)$ where $X$, $Y$, and $Z$ are jointly normal. To shows its identity, we follow a similar approach as before.
\begin{align*}
    \mathbb{E}(XYZ) =& \mathbb{E}_{Y,Z}\{\mathbb{E}_{X|Y,Z}\{X y z | Y = y, Z = z\}\} \\ 
    =& \mathbb{E}_{Y,Z}\{Y Z (\mu_X + \mathbf{\Sigma}_{XZ} \mathbf{\Sigma}_{YZ}^{-1}(y-\mu_Y, z-\mu_Z)^T)\} \\
    =& \mu_X \mu_Y \mu_Z + \mu_X \sigma_{YZ} \\ &+ D^{-1} \mathbb{E}_{Y,Z}\{Y Z ((Y-\mu_Y)(\sigma_{XY}\sigma_{ZZ}-\sigma_{XZ}\sigma_{YZ}) \\ &+ (Z-\mu_Z)(\sigma_{XZ}\sigma_{YY}-\sigma_{XY}\sigma_{YZ}))\}
\end{align*}
where $D = \sigma_{YY} \sigma_{ZZ} - \sigma_{YZ}^2$. Now, using the previously shown expectation of $\mathbb{E}(X^2 Y)$ and simplifying terms, this yields:
\begin{align*}
    =& \mu_X\mu_Y\mu_Z + \mu_X\sigma_{YZ} + D^{-1}(\mu_Z \sigma_{YY} + \mu_Y\sigma_{YZ})(\sigma_{XY}\sigma_{ZZ}-\sigma_{XZ}\sigma_{YZ}) \\ &+ D^{-1}(\mu_Y \sigma_{ZZ} + \mu_Z \sigma_{YZ})(\sigma_{XZ}\sigma_{YY} - \sigma_{XY} \sigma_{YZ}) \\
    =& \mu_X\mu_Y\mu_Z + \mu_X\sigma_{YZ} + D^{-1} \mu_Y (\sigma_{XY}\sigma_{ZZ}\sigma_{YY} - \sigma_{XY}\sigma_{YZ}^2) + D^{-1}\mu_Z (\sigma_{XY}\sigma_{ZZ}\sigma_{YY}-\sigma_{XY}\sigma_{YZ}^2) \\
    =& \mu_X\mu_Y\mu_Z + \mu_X\sigma_{YZ} + \mu_Y\sigma_{XZ} + \mu_Z\sigma_{YZ}
\end{align*}

These two expectations will be used repeatedly in deriving the forward equations. We suppress the inclusion of $\Delta t$. It should be understood the probabilities are multiplied by $\Delta t$ and then a limit is taken as $\Delta t \to 0$. The forward equation for $\mu_Y(\bs_i,t)$ is:
\begin{align*}
    \frac{\partial \mu_Y(\bs_i,t)}{\partial t} =& \mathbb{E}\{((Y(\bs_i,t)+ 1) - Y(\bs_i,t))\bigl(\frac{\beta(\bs_i)}{N(\bs_i)}X(\bs_i,t)Y(\bs_i,t) \\ &+ \frac{\phi}{N(\bs_i)}\sum_{k\in\mathcal{N}(\bs_i)}X(\bs_i,t)(Y\bs_k,t)\bigr) \\ &+ \bigl((Y(\bs_i,t) - 1) - Y(\bs_i,t)\bigr)\eta Y(\bs_i, t) \} \\
    =& \mathbb{E}\{\frac{\beta(\bs_i)}{N(\bs_i)} X(\bs_i,t)Y(\bs_i,t) + \frac{\phi}{N(\bs_i)} \sum_{k \in \mathcal{N}(\bs_i)} X(\bs_i,t)Y(\bs_k,t) - \eta Y(\bs_i, t)\} \\
    =& \frac{\beta(\bs_i)}{N(\bs_i)} \bigl(\mu_X(\bs_i, t)\mu_Y(\bs_i,t) + \sigma_{XY}(\bs_i, \bs_i;t)\bigr) \\ &+ \frac{\phi}{N(\bs_i)}\sum_{k \in \mathcal{N}(\bs_i)}\bigl(\mu_X(\bs_i,t)\mu_Y(\bs_k,t)+\sigma_{XY}(\bs_i,\bs_k;t)\bigl)-\eta\mu_Y(\bs_i,t)
\end{align*}

The forward equation for $\sigma_{XX}(\bs_i, \bs_i; t)$ is:
\begin{align*}
    \frac{\partial \sigma_{XX}(\bs_i,\bs_i;t)}{\partial t} =& \mathbb{E}\{ \bigl((\Xsi - \muxsi - 1)^2 - (\Xsi - \muxsi)^2\bigr) \\ &* \bigl(\betasnsi \Xsi \Ysi + \phinsi \sum_{\skinnsi} \Xsi \Ysk \} \\
    =& \mathbb{E}\{-2\betasnsi \Xsi^2 \Ysi + 2 \betasnsi \muxsi \Xsi \Ysi \\ &+ \betasnsi \Xsi \Ysi + \phinsi \sum_{\skinnsi}\bigl(-2\Xsi^2\Ysk \\ &+ 2\muxsi\Xsi\Ysk+\Xsi\Ysk\bigr)\} \\
    =& \betasnsi\bigl(-2\muysi \sxxsisi + \muxsi \muysi + \sxysisi \\ &- 2\muxsi\sxysisi\bigr) + \phinsi \sum_{\skinnsi}\bigl(-2\muysk\sxxsisi \\ &+ \muxsi \muysk + \sxxsisk - 2\muxsi\sxysisk\bigr)
\end{align*}

The forward equation for $\sigma_{YY}(\bs_i, \bs_i; t)$ is:
\begin{align*}
    \frac{\partial \sigma_{YY}(\bs_i,\bs_i;t)}{\partial t} =& \mathbb{E}\{\bigl(\Ysi-\muysi+1)^2-(\Ysi-\muysi)^2\bigr)\\ &* \betasnsi\Xsi\Ysi \\
    &+ \bigl(\Ysi-\muysi+1)^2-(\Ysi-\muysi)^2\bigr) \\ &* \phinsi\sum_{\skinnsi}\Xsi\Ysk \\
    &+ \bigl(\Ysi-\muysi-1)^2-(\Ysi-\muysi)^2\bigr)\eta\Ysi\} \\
    =& \mathbb{E}\{\betasnsi \bigl(2 \Xsi \Ysi^2 - 2 \muysi \Xsi \Ysi \\
    &+ \Xsi\Ysi\bigr) + \phinsi \sum_{\skinnsi}\bigl(2\Xsi\Ysi\Ysk\\&-2\muysi\Xsi\Ysk+\Xsi\Ysk\bigr) \\
    &+ \eta(-2\Ysi^2 + 2\muysi\Ysi + \Ysi)\} \\
    =& \betasnsi \bigl(2 \muxsi \syysisi + 2 \muysi \sxysisi 
    \\ &+ \muxsi \muysi + \sxysisi\bigr) \\
    &+ \phinsi \sum_{\skinnsi}\bigl(2\muxsi\syysisk + 2\muysk\sxysisi \\
    &+ \muxsi\muysk + \sxysisk\bigr) + \eta(\muysi - 2\syysisi)
\end{align*}

The remaining four forward equations may be derived in exactly the same manner.

The sample moments are well approximated by the moment-closure approximation, as seen in the example plots below.
\begin{figure}[H]
\caption{Example plots of covariances using simulations from the spatial SIR jump process (dots) and the moment-closure approximation (lines). The left plot shows a sample marginal variance for the susceptibles, and the right plot shows a sample covariance between the susceptibles and infectious at non-adjacent spatial locations.}
\label{fig:examplemoments}
\begin{center}
\begin{tabular}{cc}
\includegraphics[width=0.45\textwidth]{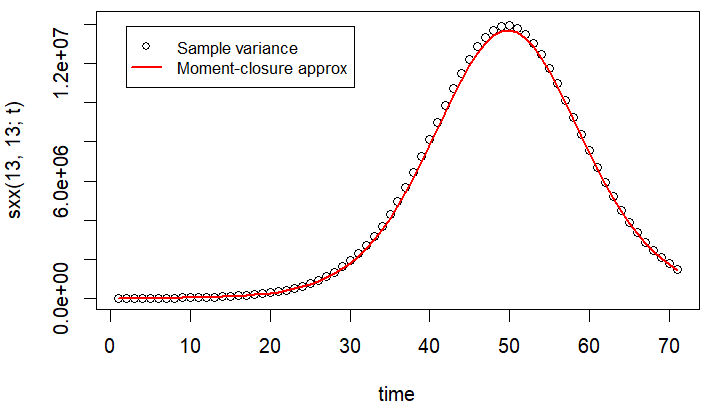} &
\includegraphics[width=0.45\textwidth]{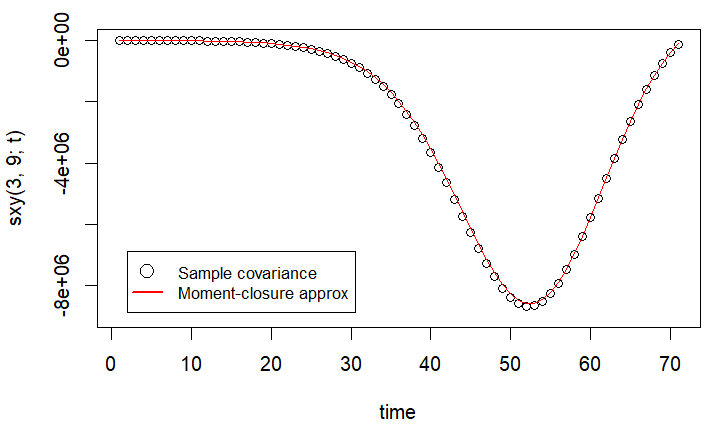}
\end{tabular}
\end{center}
\end{figure}

\subsection{Deriving the forward equations using jump-process approximation}
\label{app:fe_whittle}

We now demonstrate how to derive the forward equations using a normal approximation to the jump process as described in \cite{whittle1957} and used in \cite{isham1991}. We begin by reviewing the basic setup as discussed in \cite{whittle1957} and demonstrate its use in calculating the forward equations from \cite{isham1991}. We then provide a thorough overview on how the same approach can be used for our spatial SIR jump process.

\subsubsection{Example using stochastic SIR process approximation}

Define $\phi(\pmb{\theta};t)=\mathbb{E}_{\pmb{X}(t)}(e^{\theta^T \pmb{X}(t)})$ as the moment-generating function (mgf) of a \newline continuous-time Markov process at time $t$. A relation\footnote{See \cite{bartlett1949} and \cite{bailey1964} Section 7.4} that \cite{whittle1957} refers to as the Bartlett relation is:

\begin{align}
    \frac{\partial \phi(\pmb{\theta};t)}{\partial t} =& L(\pmb{\theta}, \frac{\partial}{\partial \pmb{\theta}};t) \phi(\pmb{\theta};t).
\end{align}

The function $L(\pmb{\theta},\frac{\partial}{\partial \pmb{\theta}};t)$ is of particular importance in this approximation method. \cite{whittle1957} calls this function the ``derivate cumulant function", and it arises when taking the derivative of the mgf with respect to time. We follow \cite{bailey1964} Section 7.4 who shows how to derive this function for a continuous-time Markov process with two or more variables. Its form is a relatively simple function of the transition probabilities.

Define $\psi(\pmb{\theta};t)=\log(\phi(\pmb{\theta};t))$ as the cumulant-generating function, $\Psi(\pmb{\theta};t)$ as the derivative of $\psi(\pmb{\theta};t)$ with respect to $\pmb{\theta}$, and $G(\pmb{\theta},\pmb{\xi};t)$ as the collection of second-order and higher terms of the Taylor expansion of $\psi(\pmb{\theta} + \pmb{\xi};t)$ around $\psi(\pmb{\theta};t)$. \cite{whittle1957} shows the Bartlett relation can be rewritten as:
\begin{align}
    \frac{\partial \psi(\pmb{\theta};t)}{\partial t} =& e^{G(\pmb{\theta}, \frac{\partial}{\partial \Psi(\pmb{\theta})};t)} L(\pmb{\theta},\Psi(\pmb{\theta};t);t)
\end{align}
The normal approximation is then complete by using the cumulant-generating function of a multivariate normal random variable at time $t$ with mean $\pmb{\mu}(t)$ and covariance $\mathbf{\Sigma}(t)$. Using $\psi(\pmb{\theta};t) = \pmb{\theta}^T \pmb{\mu}(t) + \frac{1}{2}\pmb{\theta}^T \mathbf{\Sigma}(t) \pmb{\theta}$, this yields:
\begin{align}
    \frac{\partial}{\partial t}(\pmb{\theta}^T \pmb{\mu}(t) + \frac{1}{2}\pmb{\theta}^T \mathbf{\Sigma}(t) \pmb{\theta}) =& \exp\{\frac{1}{2}(\frac{\partial}{\partial \pmb{\mu}(t)})^T \mathbf{\Sigma}(t) (\frac{\partial}{\partial \pmb{\mu}(t)})\} L(\pmb{\theta},\pmb{\mu}(t) + \mathbf{\Sigma}(t)\pmb{\theta};t)\label{eq:whittle}.
\end{align}

Though this is hard to interpret and derive, using it is nevertheless relatively straightforward. To derive the forward equations, powers of $\pmb{\theta}$ need to be matched between the left- and right-hand sides. To do so, it is necessary to use Maclaurin-series expansions in a few places. As an illustrative example, consider the function $L(\pmb{\theta}, \pmb{x};t)$ used in \cite{isham1991}:
\begin{align}
    L(\pmb{\theta},\pmb{x};t) =& \beta (e^{\theta_2 - \theta_1} - 1)x_1 x_2 / n + \eta (e^{-\theta_2} - 1) x_2 \label{eq:ishamL}
\end{align}
Begin by rewriting \eqref{eq:ishamL} using the first three terms of the Maclaurin expansions of $e^{\theta_2-\theta_1}$ and $e^{-\theta_2}$, which we will refer to as $\tilde{L}(\pmb{\theta},\pmb{x};t)$. This yields:
\begin{align}
    \tilde{L}(\pmb{\theta},\pmb{x};t) =& \beta (\theta_2 - \theta_1 + \frac{1}{2}(\theta_2 - \theta_1)^2)x_1 x_2 / n + \eta (-\theta_2 + \frac{1}{2}\theta_2^2)x_2
\end{align}

Referring back to \eqref{eq:whittle}, define $K = \frac{1}{2}(\frac{\partial}{\partial \pmb{\mu(t)}})^T \mathbf{\Sigma}(t) (\frac{\partial}{\partial \pmb{\mu}(t)})$. For readability, we will begin to suppress the $t$ notation for the mean and covariance parameters. Using the first two terms of the Maclaurin expansion of $e^K$ and substituting $\tilde{L}(\pmb{\theta},\pmb{x};t)$ for $L(\pmb{\theta},\pmb{x};t)$, rewrite \eqref{eq:whittle} as:
\begin{align}
    \frac{\partial}{\partial t}(\pmb{\theta}^T \pmb{\mu} + \frac{1}{2}\pmb{\theta}^T \mathbf{\Sigma} \pmb{\theta}) =& (1 + K)\tilde{L}(\pmb{\theta},\pmb{\mu}+\mathbf{\Sigma}\pmb{\theta};t)
\end{align}
Note that in the above equation only the first two terms of the Maclaurin expansion of $e^K$ were needed. The $\frac{1}{2}K^2 \tilde{L}(\pmb{\theta},\pmb{\mu}+\mathbf{\Sigma}\pmb{\theta})$ and higher terms will evaluate to zero when taking the higher-order derivatives with respect to the mean parameters.

Expanding the left-hand side, we get:
\begin{align}
    \frac{\partial}{\partial t}(\theta_1 \mu_X + \theta_2 \mu_Y + \frac{1}{2}\theta_1^2 \sigma_{XX} + \theta_1\theta_2\sigma_{XY} + \frac{1}{2}\theta_2^2 \sigma_{YY})\label{eq:IshamLHS}
\end{align}
The next step is to simplify the right-hand side. It is easiest to begin with the term $K \tilde{L}(\pmb{\theta},\pmb{\mu}+\mathbf{\Sigma}\pmb{\theta};t)$ because so many terms drop out. Substituting back in for $K$ and expanding slightly, it takes the form:
\begin{align}
    \frac{1}{2}(\sigma_{XX}\frac{\partial^2}{\partial \mu_X^2} + \sigma_{YY}\frac{\partial^2}{\partial \mu_Y^2} + 2\sigma_{XY}\frac{\partial}{\partial\mu_X}\frac{\partial}{\partial \mu_Y})\tilde{L}(\pmb{\theta},\pmb{\mu}+\mathbf{\Sigma}\pmb{\theta})\label{eq:solvestep1}
\end{align}
Now observe the form of $\tilde{L}(\pmb{\theta},\pmb{\mu}+\mathbf{\Sigma}\pmb{\theta};t)$:
\begin{align}
\begin{split}
   \beta (\theta_2 - \theta_1 + \frac{1}{2}(\theta_2 - \theta_1)^2)&(\mu_X + \theta_1 \sigma_{XX} + \theta_2 \sigma_{XY}) (\mu_Y + \theta_1 \sigma_{XY} + \theta_2 \sigma_{YY}) / n \\
   &+ \eta (-\theta_2 + \frac{1}{2}\theta_2^2)(\mu_Y + \theta_1 \sigma_{XY} + \theta_2 \sigma_{YY})\label{eq:expandedIshamLtilde}
\end{split}
\end{align}
There are no terms in \eqref{eq:expandedIshamLtilde} that contain either $\mu_X^2$ or $\mu_Y^2$, so the first two differential terms in \eqref{eq:solvestep1} will evaluate to zero. Observe next that there is only a single term that will contain both $\mu_X$ and $\mu_Y$, so we can quickly conclude that:
\begin{align}
    K \tilde{L}(\pmb{\theta},\pmb{\mu}+\mathbf{\Sigma}\pmb{\theta};t)&=(\theta_2-\theta_1+\frac{1}{2}(\theta_2-\theta_1)^2)\beta\sigma_{XY}/n\label{eq:simpleKL}
\end{align}

The right-hand side is therefore:
\begin{align}
\tilde{L}(\pmb{\theta},\pmb{\mu}+\mathbf{\Sigma}\pmb{\theta};t) + (\theta_2-\theta_1+\frac{1}{2}(\theta_2-\theta_1)^2)\beta\sigma_{XY}/n
\end{align}

All that remains is algebra and collecting terms for $\theta_1$ and $\theta_2$. There will be terms on the right-hand side with terms like $\theta_1^2 \theta_2^2$ and $\theta_1^4$, but those are discarded as part of the approximation. We instead need to collect the coefficients for the $\theta_1$, $\theta_2$, $\theta_1^2$, $\theta_2^2$, and $\theta_1 \theta_2$ terms. Doing so on the right-hand side will give us:
\begin{align}
    \to& \theta_1 (-\beta \mu_X \mu_Y / N) + \theta_2 (-\eta \mu_Y + \beta \mu_X \mu_Y / N) \\
    &+ \theta_1^2 (\beta/n) (-\sigma_{XY}\mu_X -\sigma_{XX}\mu_Y + \mu_X \mu_Y / 2 + \sigma_{XY}) \\
    &+ \theta_2^2 \bigl((\beta/n)(\sigma_{YY}\mu_X + \sigma_{XY}\mu_Y + \mu_X\mu_Y/2 + \sigma_{XY}/2) + \eta(-\sigma_{YY}+\mu_Y/2)\bigr) \\
    &+ \theta_1 \theta_2 \bigl((\beta/n)(\sigma_{XY}\mu_X-\sigma_{YY}\mu_X + \sigma_{XX}\mu_Y-\sigma_{XY}\mu_Y-\mu_X\mu_Y -\sigma_{XY})-\eta\sigma_{XY}\bigr)
\end{align}

Adjusting for the $1/2$ coefficients on $\theta_1^2$ and $\theta_2^2$ on the left-hand side yields the forward equations from \cite{isham1991}.

\subsubsection{Deriving the spatial SIR process forward equations using the Whittle approximation}
We now show how to arrive at the forward equations using the \cite{whittle1957} approximation.

First, we must derive $L(\pmb{\theta}, \pmb{x};t)$ for our process. As stated before, \cite{bailey1964} Section 7.4 provides details and a proof. Using his work, it is possible to immediately arrive at $L(\pmb{\theta}, \pmb{x};t)$ for our process. We adjust the notation in this part from the rest of the dissertation to improve readability. Suppose we have $n$ spatial sites, and for a spatial site $j$ there is a collection of neighbors $\mathcal{N}(j)$ that does not include $j$ itself. $\pmb{\theta}\in \mathbb{R}^{2n}$, but for simplicity, index $\pmb{\theta}$ using the convention $j,l$ where $j \in \{1,...,n\}$ corresponds to the spatial site and $l \in \{1,2\}$ corresponds to the terms for $X(s,t)$ and $Y(s,t)$, the number of susceptibles and infectious at site $s$ and time $t$, respectively. We use $\beta_j$, $\phi$, and $\eta$ for the local infection at site $j$, spatial infection, and recovery parameters as before. $N_j$ is the population at site $j$ and is assumed to be constant through time. Then the $L(\pmb{\theta},\pmb{x};t)$ for our jump process is:
\begin{align}
\begin{split}
    L(\pmb{\theta},\pmb{x};t) =& \sum_{j=1}^n (e^{-\theta_{j1}+\theta_{j2}}-1)\frac{\beta_j}{N_j}x_{j1}x_{j2} \\
    +& \sum_{j=1}^n \sum_{k \in \mathcal{N}(j)} (e^{-\theta_{j1}+\theta_{j2}}-1)\frac{\phi}{N_j}x_{j1}x_{k2} \\
    +& \sum_{j=1}^n (e^{-\theta_{j2}}-1)\eta x_{j2} \label{eq:ourL}
\end{split}
\end{align}
As before, define $\tilde{L}(\pmb{\theta},\frac{\partial}{\partial \pmb{\theta}};t)$ as above but with replacing the exponentials with the first three terms of their Maclaurin expansions. In other words:
\begin{align}
\begin{split}
    \tilde{L}(\pmb{\theta},\pmb{x};t)=& \sum_{j=1}^n (\theta_{j2}-\theta_{j1}+\frac{1}{2}(\theta_{j2}-\theta_{j1})^2)\frac{\beta_j}{N_j}x_{j1}x_{j2} \\
    +& \sum_{j=1}^n \sum_{k \in \mathcal{N}(j)} (\theta_{j2}-\theta_{j1}+\frac{1}{2}(\theta_{j2}-\theta_{j1})^2)\frac{\phi}{N_j}x_{j1}x_{k2} \\
    +& \sum_{j=1}^n (-\theta_{j2}+\frac{1}{2}\theta_{j2}^2)\eta\, x_{j2}
\end{split}
\end{align}

Stack $\pmb{\theta} = (\theta_{1,1},...\theta_{n,1},\theta_{1,2},...\theta_{n,2})^T$, and likewise consider the process at time $t$ to be stacked as $(X(1,t),...,X(n,t),Y(1,t),...,Y(n,t))^T$ with mean and covariance parameters correspondingly defined. Define again $K = \frac{1}{2}(\frac{\partial}{\partial \pmb{\mu}})^T \mathbf{\Sigma} (\frac{\partial}{\partial \pmb{\mu}})$. It then remains as before to match terms for $\pmb{\theta}$ on the left- and right-hand sides of:
\begin{align}
\begin{split}
    \frac{\partial}{\partial t}(\pmb{\theta}^T \pmb{\mu} + \frac{1}{2}\pmb{\theta}^T \mathbf{\Sigma} \pmb{\theta}) =& (1 + K)\tilde{L}(\pmb{\theta},\pmb{\mu}+\mathbf{\Sigma}\pmb{\theta};t) \label{eq:ugly}
\end{split}
\end{align}
The algebra is simplified greatly by careful accounting. Beginning with the $K \tilde{L}(\pmb{\theta},\pmb{\mu}+\mathbf{\Sigma}\pmb{\theta};t)$ term, notice that no terms in $\tilde{L}(\pmb{\theta},\pmb{\mu}+\mathbf{\Sigma}\pmb{\theta};t)$ will contain any squared terms for any particular element of $\pmb{\mu}$, so all terms with a squared differential operator will drop out. Furthermore, there are no cross-terms in the final summation (i.e., in the summation containing $\eta$, $x_{j2}$ is not multiplied by any other element of $\pmb{x}$), which will lead these terms to also dropping out when taking the derivative with respect to two different elements of $\pmb{\mu}$. Close inspection then shows that most of the remaining terms also drop out -- for example, there are no terms that contain both $\mu_X(j)$ and $\mu_X(k)$ with $j \neq k$, so the $(\frac{\partial}{\partial \mu_X(j)})(\frac{\partial}{\partial\mu_X(k)})\tilde{L}(\pmb{\theta},\pmb{\mu}+\mathbf{\Sigma}\pmb{\theta};t)$ terms will not appear. That applies similarly for $\mu_X(j)$ and $\mu_Y(k)$ and so on. What we find is all that will remain is:
\begin{align}
\begin{split}
    K \tilde{L}(\pmb{\theta},\pmb{\mu}+\mathbf{\Sigma}\pmb{\theta};t) =& \sum_{j=1}^n \sigma_{xy}(j,j) \frac{\beta_j}{N_j}(\theta_{j2}-\theta_{j1}+\frac{1}{2}(\theta_{j2}-\theta_{j1})^2) \\
    +& \sum_{j=1}^n \sum_{k\in\mathcal{N}(j)} \sigma_{xy}(j,k) \frac{\phi}{N_j} (\theta_{j2}-\theta_{j1}+\frac{1}{2}(\theta_{j2}-\theta_{j1})^2) \label{eq:nicer}
\end{split}
\end{align}
Therefore, to arrive at the forward equations, we track the coefficients of certain powers and interactions of the elements of $\pmb{\theta}$ in $\tilde{L}(\pmb{\theta},\pmb{\mu}+\mathbf{\Sigma}\pmb{\theta};t)$ and \eqref{eq:nicer}. For an arbitrary $j$, collect the coefficients for $\theta_{j,1}$ excluding any interactions with other elements of $\pmb{\theta}$. This leads to the forward equation:
\begin{align}
    \frac{\partial \mu_X(j)}{\partial t} = -\frac{\beta_j}{N_j}(\mu_X(j) \mu_Y(j) + \sigma_{XY}(j,j)) - \sum_{k\in\mathcal{N}(j)}\frac{\phi}{N_j}(\mu_X(j)\mu_Y(k) + \sigma_{XY}(j,k))
\end{align}
Likewise for an arbitrary $j$, collect the terms that contain $\theta_{j,2}$. This leads to:
\begin{align}
    \frac{\partial \mu_Y(j)}{\partial t} = \frac{\beta_j}{N_j}(\mu_X(j) \mu_Y(j) + \sigma_{XY}(j,j)) + \sum_{k\in\mathcal{N}(j)}\frac{\phi}{N_j}(\mu_X(j)\mu_Y(k) + \sigma_{XY}(j,k)) - \eta \mu_Y(j)
\end{align}
For an arbitrary $j$, the forward equations for $\sigma_{XX}(j,j)$, $\sigma_{YY}(j,j)$, and $\sigma_{XY}(j,j)$ can be found by collecting the terms that contain $\theta_{j1}^2$, $\theta_{j2}^2$, and $\theta_{j1} \theta_{j2}$, respectively. These yield:
\begin{align}
\begin{split}
    \frac{\partial \sigma_{XX}(j,j)}{\partial t} =& \frac{\beta_j}{N_j}(\mu_X(j)\mu_Y(j)+\sigma_{XY}(j,j)-2\mu_X(j)\sigma_{XY}(j,j)-2\mu_Y(j)\sigma_{XX}(j,j)) \\
    +& \sum_{k\in\mathcal{N}(j)}\frac{\phi}{N_j}(\mu_X(j)\mu_Y(k)+\sigma_{XY}(j,k) - 2 \mu_Y(k)\sigma_{XX}(j,j) - 2 \mu_X(j) \sigma_{XY}(j,k)) \\
\end{split}
\end{align}
\begin{align}
\begin{split}
    \frac{\partial \sigma_{YY}(j,j)}{\partial t} =& \frac{\beta_j}{N_j}(\mu_X(j)\mu_Y(j)+\sigma_{XY}(j,j) + 2\mu_Y(j)\sigma_{XY}(j,j) + 2\mu_X(j)\sigma_{YY}(j,j)) \\
    +& \sum_{k\in\mathcal{N}(j)}\frac{\phi}{N_j}(\mu_X(j)\mu_Y(k) + \sigma_{XY}(j,k) + 2\mu_X(j)\sigma_{YY}(j,k)+2\mu_Y(k)\sigma_{XY}(j,j)) \\
    +& \eta(\mu_Y(j) - 2 \sigma_{YY}(j,j))
\end{split}
\end{align}
\begin{align}
\begin{split}
    \frac{\partial \sigma_{XY}(j,j)}{\partial t} =& \frac{\beta_j}{N_j}(-\mu_X(j)\mu_Y(j)-\sigma_{XY}(j,j)-\mu_X(j)\sigma_{YY}(j,j)-\mu_Y(j)\sigma_{XY}(j,j) \\
    +&\mu_X(j)\sigma_{XY}(j,j) + \mu_Y(j)\sigma_{XX}(j,j)) + \sum_{k\in\mathcal{N}(j)} \frac{\phi}{N_j}(-\mu_X(j)\mu_Y(k)-\sigma_{XY}(j,k)\\
    -& \mu_X(j)\sigma_{YY}(j,k)-\mu_Y(k)\sigma_{XY}(j,j)+\mu_X(j)\sigma_{XY}(j,k)+\mu_Y(k)\sigma_{XX}(j,j)) \\
    -&\eta\sigma_{XY}(j,j)
\end{split}
\end{align}
Given $j\neq k$, forward equations can be derived for $\frac{\partial \sigma_{XX}(j,k)}{\partial t}$ and $\frac{\partial \sigma_{YY}(j,k)}{\partial t}$ by finding terms with $\theta_{j1}\theta_{k1}$ and $\theta_{j2}\theta_{k2}$, respectively. $\frac{\partial \sigma_{XY}(j,k)}{\partial t}$ can be derived by finding terms with $\theta_{j1}\theta_{k2}$.
\begin{align}
\begin{split}
    \frac{\partial \sigma_{XX}(j,k)}{\partial t} =& -\frac{\beta_j}{N_j}(\mu_X(j)\sigma_{XY}(k,j)+\mu_Y(j)\sigma_{XX}(j,k)) \\-& \frac{\beta_k}{N_k}(\mu_X(k)\sigma_{XY}(k,j)+\mu_Y(k)\sigma_{XX}(j,k)) \\
    +& \sum_{l\in\mathcal{N}(j)}\frac{\phi}{N_j}(-\mu_X(j)\sigma_{XY}(k,l)-\mu_Y(l)\sigma_{XX}(j,k)) \\
    +& \sum_{m\in\mathcal{N}(k)}\frac{\phi}{N_k}(-\mu_X(k)\sigma_{XY}(j,m)-\mu_Y(m)\sigma_{XX}(j,k))
\end{split}
\end{align}
\begin{align}
\begin{split}
    \frac{\partial \sigma_{YY}(j,k)}{\partial t} =& \frac{\beta_j}{N_j}(\mu_X(j)\sigma_{YY}(j,k)+\mu_Y(j)\sigma_{XY}(j,k))\\+&\frac{\beta_k}{N_k}(\mu_X(k)\sigma_{YY}(j,k)+\mu_Y(k)\sigma_{XY}(k,j)) \\
    +& \sum_{l\in\mathcal{N}(j)}\frac{\phi}{N_j}(\mu_X(j)\sigma_{YY}(k,l) + \mu_Y(l)\sigma_{XY}(j,k)) \\
    +& \sum_{m\in\mathcal{N}(k)}\frac{\phi}{N_k}(\mu_X(k)\sigma_{YY}(j,m) + \mu_Y(m)\sigma_{XY}(k,j)) \\
    -& 2\eta\sigma_{YY}(j,k)
\end{split}
\end{align}
\begin{align}
\begin{split}
    \frac{\partial \sigma_{XY}(j,k)}{\partial t} =& \frac{\beta_j}{N_j}(-\mu_X(j)\sigma_{YY}(j,k)-\mu_Y(j)\sigma_{XY}(j,k)) \\
    +& \frac{\beta_k}{N_k}(\mu_X(k)\sigma_{XY}(j,k)+\mu_Y(k)\sigma_{XX}(j,k)) \\
    +& \sum_{l\in\mathcal{N}(j)}\frac{\phi}{N_j}(-\mu_X(j)\sigma_{YY}(k,l)-\mu_Y(l)\sigma_{XY}(j,k)) \\
    +& \sum_{m\in\mathcal{N}(k)}\frac{\phi}{N_k}(\mu_X(k)\sigma_{XY}(j,m)+\mu_Y(m)\sigma_{XX}(j,k))\\
    -&\eta(2\sigma_{XY}(j,k))
\end{split}
\end{align}


\section{Overview of tensor mathematics}
\label{app:tensors}
We summarize the tensor math we use to construct our emulators. For a more thorough introduction to tensors and tensor decompositions, see \cite{koldabader}. 

A tensor is an array of numbers with $N$ orders, where the number of orders is the number of indices (also called the number of ``modes''). A first-order tensor is a vector, a second-order tensor is a matrix, and a third-order tensor may be envisioned as a rectangular prism of numbers. For simplicity, we only call an array a tensor if it has at least three modes. We adopt the common convention to use calligraphic font for tensors.

Our emulator basis-function algorithms are based on the higher-order singular value decompositions (HOSVD) of tensors \citep{hosvdpaper, koldabader}. A HOSVD is an extension of the singular value decomposition (SVD) of matrices to tensors. It decomposes an $N$th-order tensor into $N$ factor matrices with orthonormal columns and a resulting core tensor $\mathcal{Z}$. These factor matrices may be thought of as the tensor equivalent of the matrices containing the left and right singular vectors in an SVD, or alternatively as analogues to loadings in principal component analysis \citep{islr2e} or empirical orthogonal functions in spatial analysis \citep{cressiewikle}. As with a low-rank approximation to a matrix based on an SVD, the goal is for the factor matrices to serve as orthogonal basis functions that capture most of the variability along the tensor's modes when measured by a Frobenius norm. The columns are orthonormal, and they may be considered as estimates for the eigenvectors of the uncentered second moments of the spatial and temporal processes, i.e., $\text{Cov}\{\bmu_X(t,\btheta)\}+\mathbb{E}\{\bmu_X(t,\btheta)\}\mathbb{E}\{\bmu_X(t,\btheta)\}^T$.

An HOSVD for a third-order tensor $\mathcal{A} \in \mathbb{R}^{a \times b \times c}$ may be written as
\begin{equation}
    \mathcal{A} = \mathcal{Z} \times_1 \mathbf{U}_1 \times_2 \mathbf{U}_2 \times_3 \mathbf{U}_3 \label{eq:tensor_hosvd_basic}
\end{equation}
where $\mathcal{Z}$ is the core tensor $\in \mathbb{R}^{a \times b \times c}$, $\mathbf{U}_1$ is the factor matrix for the first order $\in \mathbb{R}^{a \times a}$, $\mathbf{U}_2$ is the factor matrix for the second order $\in \mathbb{R}^{b \times b}$, and $\mathbf{U}_3$ is the factor matrix for the third order $\in \mathbb{R}^{c \times c}$. In general, the core tensor $\mathcal{Z}$ is dense, unlike its counterpart in a matrix SVD.

Symbols like $\times_1$ in \eqref{eq:tensor_hosvd_basic} indicate an ``$n$-mode product''. These tensor-matrix products involve holding indices for the non-$i$th modes fixed and then premultiplying the resulting vectors (``fibers'') by the specified matrix. For example, $\mathcal{Z} \times_1 \mathbf{U}_1$ in \eqref{eq:tensor_hosvd_basic} may be written as:

\vspace{8mm}
\begin{algorithm}[H]
\begin{singlespace}
\caption{$n$-mode multiplication example}\label{alg:tensor_nmode}
\KwData{$\mathcal{Z} \in \mathbb{R}^{a \times b \times c}$, $\mathbf{U}_1 \in \mathbb{R}^{a \times a}$}
\KwResult{$\mathcal{Y} = \mathcal{Z} \times_1 \mathbf{U}_1$}
initialize $\mathcal{Y} \in \mathbb{R}^{a \times b \times c}$\;
\For{$j \in \{1,...,b\}$}{
\For{$k \in \{1,...,c\}$} {
$\mathcal{Y}_{.,j,k} \gets \mathbf{U}_1 \mathcal{Z}_{.,j,k}$\;
}
}
\end{singlespace}
\end{algorithm}
\vspace{8mm}

We also refer to ``unfolding'' tensors. This indicates turning a tensor into a matrix, where the tensor is unfolded along a particular mode. For example, unfolding $\mathcal{A}$ along its first mode would result in a matrix $\in \mathbb{R}^{a \times bc}$. This matrix $\mathbf{K}$ may be computed as:

\vspace{8mm}
\begin{algorithm}[H]
\begin{singlespace}
\caption{Tensor unfolding example}\label{alg:tensor_unfold}
\KwData{$\mathcal{A} \in \mathbb{R}^{a \times b \times c}$}
\KwResult{$\mathbf{K} \in \mathbb{R}^{a \times bc}$, the matrix resulting from unfolding $\mathcal{A}$ along its first mode}
initialize $\mathbf{K} \in \mathbb{R}^{a \times bc}$\;
$i \gets 1$\;
\For{$k \in \{1,...,c\}$}{
\For{$j \in \{1,...,b\}$} {
$\mathbf{K}_{.,i} \gets \mathcal{A}_{.,j,k}$\;
$i \gets i + 1$\;
}
}
\end{singlespace}
\end{algorithm}
\vspace{8mm}

Using the concepts of $n$-mode multiplication and tensor unfolding, the calculation of an HOSVD is straightforward. To compute the factor matrices, unfold the tensor along each of its modes, and then calculate the left singular vectors of the SVD of the resulting matrices. The core tensor is then calculated by taking the $n$-mode products of the transposes of each of the factor matrices.

\vspace{8mm}
\begin{algorithm}[H]
\begin{singlespace}
\caption{Calculating the HOSVD}\label{alg:tensor_hosvd}
\KwData{$\mathcal{A} \in \mathbb{R}^{a \times b \times c}$}
\KwResult{A core tensor $\mathcal{Z} \in \mathbb{R}^{a\times b \times c}$; factor matrices $\mathbf{U}_1 \in \mathbb{R}^{a \times a}, \mathbf{U}_2 \in \mathbb{R}^{b \times b}, \mathbf{U}_3 \in \mathbb{R}^{c \times c}$ }
\For{$i \in \{1,2,3\}$}{
$\mathbf{K}_i \gets$ unfold $\mathcal{A}$ along mode $i$\;
$\mathbf{U}_i \gets$ matrix of left-singular vectors of SVD of $\mathbf{K}_i$\;
}
$\mathcal{Z} \gets \mathcal{A} \times_1 \mathbf{U}_1^T \times_2 \mathbf{U}_2^T \times_3 \mathbf{U}_3^T$\;
\end{singlespace}
\end{algorithm}
\vspace{8mm}

To calculate a low-rank HOSVD, use only the first $r_i$ left singular vectors of $\mathbf{K}_i$ in Algorithm \ref{alg:tensor_hosvd}. The resulting core tensor is then $\mathcal{Z} \in \mathbb{R}^{r_1 \times r_2 \times r_3}$, where $r_1 \leq a$, $r_2 \leq b$, and $r_3 \leq c$. It follows naturally that $\mathbf{U}_1 \in \mathbb{R}^{a \times r_1}$, $\mathbf{U}_2 \in \mathbb{R}^{b \times r_2}$, and $\mathbf{U}_3 \in \mathbb{R}^{c \times r_3}$.


\section{Suggestions on implementation of full model}
\label{app:implementation}

We now provide some suggestions on how to implement our emulator-based model, such as how to tune the emulator designs and how to fit the model by MCMC.

\subsection{Implementation discussion}
\label{app:implementation_emulator}

Implementing our emulator-based approximations to $\bmu_X(\btheta)$ and $\mathbf{\Sigma}_{XX}(t,\btheta)$ requires a choice for $K$, $J_s$, $J_t$, $L_s$, and $L_t$. We now provide recommendations on choosing these values.

To choose $J_s$, $J_t$, $L_s$, and $L_t$, we suggest a similar approach to the one described in \cite{gopalanwikle}. For example with the simulated mean output stored in $\mathcal{U}$, we calculate the percentage of sums of squares explained as ${1 - \frac{|| \mathcal{U} - \hat{\mathcal{U}}||_F^2}{||\mathcal{U} - \bar{\mathcal{U}}||_F^2}}$, where $\hat{\mathcal{U}}$ is the $J_s \times J_t \times K$ rank approximation to $\mathcal{U}$, $\bar{\mathcal{U}} \in \mathbb{R}^{n_s \times n_t \times K}$ is a tensor containing the overall mean of $\mathcal{U}$, and the subscript $F$ indicates the Frobenius norm. We proceed similarly for choosing $L_s$ and $L_t$, though we typically will choose $L_s < J_s$ based on run-time considerations.

The choice of $K$ is tied to the space-filling design used for the emulator experiments. Initial estimates should be made for $\bbeta$ and $\phi$, and then experimental values for $\bbeta$ and $\phi$ should be chosen in an interval around the initial estimates. We found $\pm 25\%$ to be a good starting point. Because we use GP regression with nearest neighbors to estimate $\hat{\mathcal{m}}$ and $\hat{\mathcal{M}}$, the only cost to increasing the experimental intervals is the increased run time to generate the forward equations and calculate the resulting basis functions and weights for the emulators. The choice of $K$ then is a function of the width chosen and available computational resources.

After picking $J_s$, $J_t$, $L_s$, $L_t$, and $K$, we found using five-fold cross validation to evaluate our emulation model to be insightful. Poor out-of-sample fits can be identified both by plotting the mean and covariance curves as well as comparing MSEs with different combinations of the five values.

\subsection{MCMC discussion}
\label{app:implementation_mcmc}

We found using naive Metropolis Hastings updates could lead to poor posterior convergence. In particular, the chains for $\bbeta$ and $\phi$ will usually be negatively correlated because an increase in local infections can be offset by a decrease in spatial infections (and vice versa). However, updating $\bbeta$ and $\phi$ together but separately from $\balpha$ can still lead to the chains becoming stuck. This is because $\balpha$ captures deviations from the mean curves. Our solution was to propose $\bbeta$, $\phi$, and $\balpha_1$ (corresponding to $\mathbf{\Phi}_1$) together in a two-stage DRAM update \citep{haariolaine_dram, smith_uq}. The remaining elements of $\balpha$ are updated one-at-a-time using Metropolis Hastings updates.

Even with proper tuning of the MCMC proposals, the chains for $\bbeta$ and $\phi$ can become stuck if proposals are made outside of the parameter subspace used in the space-filling design described in Appendix \ref{app:implementation_emulator}. This can be detected easily by examining trace plots of the posterior chains. Assuming starting values for $\btheta$ are within the space-filling design, then a wider subspace should be used for the emulator design. This wider space-filling design may also necessitate a larger $K$.


\section{Algorithms and streaming calculations for emulator methodology}
\label{app:algorithms}

The low-rank approximations described in Sections \ref{sec:lowrank_mean} and \ref{sec:lowrank_cov} are based on calculating the HOSVD of a tensor. However, this necessitates storing the results for all $\bmu_X(\btheta_k)$ in a third-order tensor $\mathcal{U}$ and the results for all $\mathbf{\Sigma}_{XX}(\btheta_k)$ in a fourth-order tensor $\mathcal{S}$. Storing these tensors in memory is only possible for sufficiently small $n_s$, $n_t$, and $K$. In practice, it is unlikely these tensors can be held in memory as $n_s$, $n_t$, and $K$ collectively increase. Therefore, we implement streaming calculations for the factor matrices for our mean and covariance emulators, i.e., $\bgamma$ and $\bdelta$ for the mean emulator and $\bGamma$ and $\bDelta$ for the covariance emulator. We also calculate the respective weight tensors $\mathcal{m}$ and $\mathcal{M}$ via a streaming algorithm. This streaming approach requires either computing the forward equations twice for each $\btheta_k$ or saving the results after the first pass. The first pass is used to calculate the basis functions, and the second pass is used to calculate the weights.

\vspace{8mm}
\begin{algorithm}[H]
\begin{singlespace}
\caption{Mean emulator: Estimating basis functions and weights}\label{alg:mux_offline}
\KwData{The simulated matrices $\bmu_X(\btheta_k) \in \mathbb{R}^{n_s \times n_t}$ for $k \in \{1,...,K\}$}
\KwResult{$\bgamma = \left[\bgamma_1, ..., \bgamma_{J_s}\right]$, the spatial factor matrix; $\bdelta = \left[\bdelta_1, ..., \bdelta_{J_t}\right]$, the temporal factor matrix; $\mathcal{m}$, a tensor of weights for the interactions of $\bgamma$ and $\bdelta$}
initialize $\mathbf{\hat{B}}_S$, a $n_s \times n_s$ matrix of zeroes\;
initialize $\mathbf{\hat{B}}_T$, a $n_t \times n_t$ matrix of zeroes\;
\tcp{first pass of forward equations}
\For{$k \in \{1,...,K\}$} {
  load/run $\bmu_X(\btheta_k)$\;
  $\mathbf{\hat{B}}_S \gets \mathbf{\hat{B}}_S + \bmu_X(\btheta_k) \bmu_X(\btheta_k)^T$\;
  $\mathbf{\hat{B}}_T \gets \mathbf{\hat{B}}_T + \bmu_X(\btheta_k)^T \bmu_X(\btheta_k)$\;
}
$\bgamma \gets$ first $J_s$ eigenvectors of $\mathbf{\hat{B}}_S$\;
$\bdelta \gets$ first $J_t$ eigenvectors of $\mathbf{\hat{B}}_T$\;
initialize $\mathcal{m}$, a $J_s \times J_t \times K$ array of zeroes\;
\tcp{second pass of forward equations}
\For{$k \in \{1,...,K\}$} {
  load/run $\bmu_X(\btheta_k)$\;
  $\mathcal{m}_{.,.,k} \gets \bgamma^T \bmu_X(\btheta_k) \bdelta $\;
}
\end{singlespace}
\end{algorithm}

\vspace{8mm}

\begin{algorithm}[H]
\begin{singlespace}
\caption{Covariance emulator: Estimating basis functions and Cholesky decompositions of weights}\label{alg:sxx_offline}
\KwData{The simulated tensors $\mathbf{\Sigma}_{XX}(\btheta_k) \in \mathbb{R}^{n_s\times n_s \times n_t}$ for $k \in \{1,...,K\}$}
\KwResult{$\mathbf{\Gamma} \in \mathbb{R}^{n_s \times L_S}$, the spatial factor matrix; $\mathbf{\Delta} \in \mathbb{R}^{n_t \times L_t}$, the temporal factor matrix for the Cholesky decompositions of the weights; $\mathcal{M} \in \mathbb{R}^{L_s \times L_s \times L_t \times K}$, a tensor of the weights for the Cholesky decompositions of $\mathbf{Z}(t, \btheta_k)$}
initialize $\mathbf{A}_S$, a $n_s \times n_s$ matrix of zeroes\;
\tcp{first pass of forward equations}
\For{$k \in \{1,...,K\}$} {
  load/run $\mathbf{\Sigma}_{XX}(\btheta_k)$\;
  $\mathbf{S}_s \gets$ matrix resulting from unfolding $\mathbf{\Sigma}_{XX}(\btheta)$ along its first mode, $\in \mathbb{R}^{n_s \times (n_s n_t)}$\;
  $\mathbf{A}_S \gets \mathbf{A}_S + \mathbf{S}_s \mathbf{S}_s^T$\;
}
$\mathbf{\Gamma} \gets$ first $L_s$ eigenvectors of $\mathbf{A}_S$\;
initialize $\mathcal{C}$, a $L_s \times L_s \times n_t \times K$ array of zeroes\;
initialize $\mathbf{A}_T$, a $n_t \times n_t$ matrix of zeroes\;
\tcp{second pass of forward equations}
\For{$k \in \{1,...,K\}$} {
load/run $\mathbf{\Sigma}_{XX}(\btheta_k)$\;
$\mathcal{Z} \gets \mathbf{\Sigma}_{XX}(t,\btheta_k) \times_1 \mathbf{\Gamma}^T \times_2 \mathbf{\Gamma}^T$\;
\For{$t \in \{1,...,n_t\}$} {
$\mathcal{C}_{.,.,t,k} \gets $ Cholesky decomposition of $\mathcal{Z}_{.,.,t}$\;
}
$\mathbf{S}_T \gets$ unfolded $C_{.,.,.,k}$ along its third mode, $\in \mathbb{R}^{n_t \times J_S^2}$\;
$\mathbf{A}_T \gets \mathbf{A}_T + \mathbf{S}_T \mathbf{S}_T^T$\;
}
$\mathbf{\Delta} \gets$ first $L_t$ eigenvectors of $\mathbf{A}_T$\;
$\mathcal{M} \gets \mathcal{C} \times_3 \mathbf{\Delta}^T$\;
\end{singlespace}
\end{algorithm}
\vspace{8mm}

In the main text, we discussed our imputation methodology for unobserved weights for the mean and covariance emulators. The algorithms for those imputations are provided below.

\vspace{8mm}
\begin{algorithm}[H]
\begin{singlespace}
\caption{Mean emulator: Interpolation}\label{alg:mux_mcmc}
\KwData{$\btheta^*$, a proposed parameter vector; $\bgamma$, $\bdelta$, $\mathcal{m}$ from Algorithm \ref{alg:mux_offline}; $n_u$, the number of nearest neighbors; $\zeta_u$, a range for the Mat\'ern correlation function}
\KwResult{$\hat{\bmu}_X(\btheta^*) \in \mathbb{R}^{n_s \times n_t}$}
$\mathbf{\Theta}_R \gets n_u$ nearest neighbors to $\btheta^*$\;
$\mathbf{D} \gets$ distance among all $\mathbf{\Theta}_R$ and $\btheta^*$\;
$\mathbf{E} \gets$ Mat\'ern correlation given $\mathbf{D}$, smoothness 2.5, range $\zeta_u$\;
initialize $\mathbf{m}^*$, a $J_s \times J_t$ matrix of zeroes\;
\For{$i_1 \in \{1,...,J_s\}$} {
  \For{$i_2 \in \{1,...,J_t\}$} {
    $\mathbf{m}_{i_1, i_2}^* \gets$ ordinary kriging using $\mathcal{m}_{i_1,i_2,.}$, $\mathbf{\Theta}_R$, and $\mathbf{E}$
  }
}
$\hat{\bmu}_X(\btheta^*) \gets \bgamma \mathbf{m}^* \bdelta^T$\;
\end{singlespace}
\end{algorithm}
\vspace{8mm}

\vspace{8mm}
\begin{algorithm}[H]
\begin{singlespace}
\caption{Covariance emulator: Interpolation}\label{alg:sxx_mcmc}
\KwData{$\btheta^*$, a proposed parameter vector; $\mathbf{\Gamma}$, $\mathbf{\Delta}$, $\mathcal{M}$ from Algorithm \ref{alg:sxx_offline}; $n_c$, the number of nearest neighbors; $\zeta_c$, a range for the Mat\'ern correlation function}
\KwResult{$\mathbf{\Phi}(t,\btheta^*)$ such that $\mathbf{\Phi}(t,\btheta^*)\mathbf{\Phi}(t,\btheta^*)^T = \hat{\mathbf{\Sigma}}_{XX}(t,\btheta^*)$}
$\mathbf{\Theta}_R \gets n_c$ nearest neighbors to $\btheta^*$\;
$\mathbf{D} \gets$ Euclidean distance among all $\mathbf{\Theta}_R$ and $\btheta^*$\;
$\mathbf{E} \gets$ Mat\'ern correlation given $\mathbf{D}$, smoothness 2.5, range $\zeta_c$\;
initialize $\mathcal{M}^*$, a $L_s \times L_s \times L_t$ array of zeroes\;
\For{$j_1 \in \{1,...,L_s\}$} {
  \For{$j_2 \in \{1,..., j_1\}$} {
    \For{$j_3 \in \{1,...,L_t\}$} {
      $\mathcal{M}_{j_1, j_2, j_3}^* \gets$ ordinary kriging using $\mathcal{M}_{j_1,j_2,j_3,.}$, $\mathbf{\Theta}_R$, and $\mathbf{E}$
    }
  }
}
$\mathcal{C}^* \gets \mathcal{M}^* \times_3 \mathbf{\Delta}$\;
initialize $\mathbf{\Phi}$, a time-indexed list\;
\For{$t \in \{1,...,n_t\}$} { 
  $\mathbf{\Phi}(t) \gets \mathbf{\Gamma} \mathcal{C}_{.,.,t}^*$\;
}
\end{singlespace}
\end{algorithm}
\vspace{8mm}


\section{Proof of covariance emulator Cholesky decomposition}
\label{app:cholesky_pf}

Suppose $\mathcal{S} \in \mathbb{R}^{n_s \times n_s \times n_t \times K}$, where the slice $\mathcal{S}_{.,.,t,k}$ is a symmetric, positive definite matrix. Unfolding this tensor along its first mode is equivalent to unfolding the tensor along its first mode because of this symmetry. The unfolded tensor along the first mode, $\mathbf{S}_1$, is:
\begin{align}
    \mathbf{S}_1 &= \left[\mathcal{S}_{.,.,1,1},...,\mathcal{S}_{.,.,n_t,1},\mathcal{S}_{.,.,1,2},...,\mathcal{S}_{.,.,n_t,K}\right]\in\mathbb{R}^{n_s \times (n_s n_t K)}
\end{align}
Because $\mathcal{S}_{.,.,t,k}$ is positive definite for arbitrary $t$ and $k$, then the rank of each $\mathcal{S}_{.,.,t,k}$ is $n_s$ (i.e., the partitions of $\mathbf{S}_1$). Therefore, because the rank of each partition is at least the rank of $\mathbf{S}_1$ by \cite{harvillebook} Section 4.5, and the rank of $\mathbf{S}_1$ is at most $n_s$, the rank of $\mathbf{S}_1$ is $n_s$. Because the rank of $\mathbf{S}_1$ is $n_s$, then there are $n_s$ singular values in the SVD of $\mathbf{S}_1$, implying there are $n_s$ left singular vectors that form $\bGamma$ in the context of a full-rank HOSVD.

Now consider $\mathbf{\Sigma}_{XX}(t,\btheta) = \bGamma \mathcal{Z}_{.,.,t,k}\bGamma^T$ as in Section \ref{sec:lowrank_cov}. $\mathcal{Z}_{.,.,t,k}$ must be symmetric because $\bGamma \mathcal{Z}_{.,.,t,k} \bGamma^T = \mathbf{\Sigma}_{XX}(t,\btheta) = \mathbf{\Sigma}_{XX}(t,\btheta)^T = \bGamma \mathcal{Z}_{.,.,t,k}^T \bGamma^T$. Likewise, the rank of $\bGamma \mathcal{Z}_{.,.,t,k} \bGamma^T$ must equal the rank of $\mathbf{\Sigma}_{XX}(t,\btheta)$, which is rank $n_s$ because it is positive definite. Because $\bGamma$ is rank $n_s$, then the rank of $\mathcal{S}_{.,.,t,k}$ must be $n_s$ also. Therefore there must be a Cholesky decomposition of $\mathcal{Z}_{.,.,t,k}$.

In the case of a low-rank approximation, such that $L_s < n_s$, $\mathcal{Z}_{.,.,t,k}$ is still symmetric and positive definite. Symmetry follows from the same argument as above, and positive definiteness follows from considering that $\ba^T \bGamma^T \mathbf{\Sigma}_{XX}(t,\btheta) \bGamma \ba > 0$ when $\ba \neq \pmb{0}$, an all-zero vector. This follows from the positive definiteness of $\mathbf{\Sigma}_{XX}(t,\btheta)$. Therefore, a Cholesky decomposition will exist.


\section{Additional details on simulation studies}
\label{app:sim}

We provide additional information on the simulation studies in the main text.

\subsection{Details on sim-study 1}
\label{app:sim1}

To prepare our emulators, we construct a space-filling design over a subset of the parameter space for $\beta$ and $\phi$. We use the subspace covering the interval of $\left[0.0215, 0.0645\right]$ and $\left[0.0125, 0.0375\right]$ for $\beta$ and $\phi$, respectively. This represents a sufficiently wide subspace that knowledge of the true parameter values does not assist in estimation. We then construct a Latin hypercube design over an evenly spaced grid for these two subspaces as well as an equal number of replications of each $s \in \mathcal{D_0}$, the spatial coordinates corresponding to $S_0$ and locations $\in \mathcal{D}$ that are adjacent to $S_0$. We use $K = 10,000$.

\begin{figure}[H]
\caption{Simulated new-infection data compared with forward-equation mean and emulated mean}
\label{fig:sim_example_diag}
\begin{center}
\begin{tabular}{c}
\includegraphics[width=0.95\textwidth]{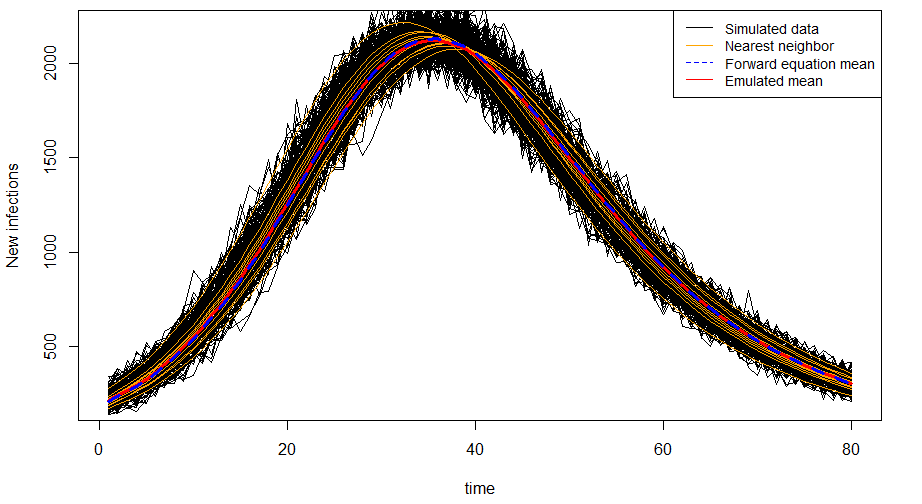}
\end{tabular}
\end{center}
\end{figure}

To evaluate the performance of our emulator design and to help choose $J_s$, $J_t$, $L_s$, and $L_t$, we examined sum-of-squares explained and analyzed out-of-sample performance using five-fold cross validation, as explained in Appendix \ref{app:implementation_emulator}. In addition, we use plots like Figure \ref{fig:sim_example_diag} to evaluate performance. This figure plots the estimated new-infection mean curve using our emulator settings against the mean curve from the forward equations and the simulated data. The orange lines are the ten nearest neighbor curves in the parameter space that are used to estimate the weights for the red curve. Results like in Figure \ref{fig:sim_example_diag} suggest good emulator performance; if the red and blue lines do not match well, it suggests needing larger $J_s$ and/or $J_t$. If the spread in the orange lines is noticeably bigger than in Figure \ref{fig:sim_example_diag}, then a larger $K$ is needed.

\vspace{-5mm}

As discussed in the main text, we tune our emulators to use $J_s = 20$, $J_t = 10$, $L_s = 10$, and $L_t = 10$. This results in more than 99.99\% of variance explained in both $\mathcal{U}$ and $\mathcal{S}$.

\subsection{Details on sim-study 2}
\label{app:sim2}

Our process for tuning the emulators is similar to the process used in the first set of simulation studies. We use a space-filling design for $\beta_0$ and $\phi$ corresponding to $\left(-3.5375, -2.1225\right)$ and $\left(0.03375, 0.05625\right)$, respectively. We needed to use a wider interval for $\beta_1$ and used $\left(-0.05, 0.25\right)$ because the chains became stuck using a narrower interval, as described in Appendix \ref{app:implementation_mcmc}. We used $K = 60,000$ to allow for the relatively wider range for $\beta_1$, and we use $J_s = 20$, $J_t = 10$, $L_s = 10$, and $L_t = 10$. This low-rank approximation again explains more than 99.99\% of variability in the the means and 72\% of the variability in the covariances.


\section{Additional details on Zika data analysis}
\label{app:zika}

We first provide an additional figure and table for the Zika data analysis in the main text.

\begin{figure}[H]
\caption{Post-burn-in trace plots of $\beta_0$ (left), $\beta_1$ (center), and $\phi$ (right).}
\label{fig:braziltraces}
\begin{center}
\begin{tabular}{ccc}
\includegraphics[width=0.3\textwidth]{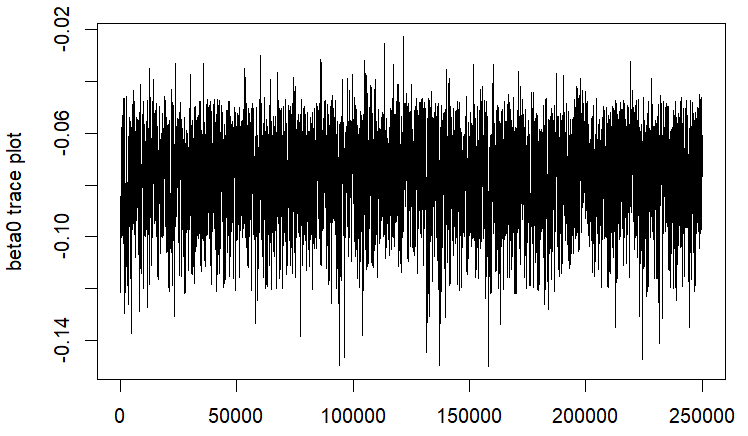} &
\includegraphics[width=0.3\textwidth]{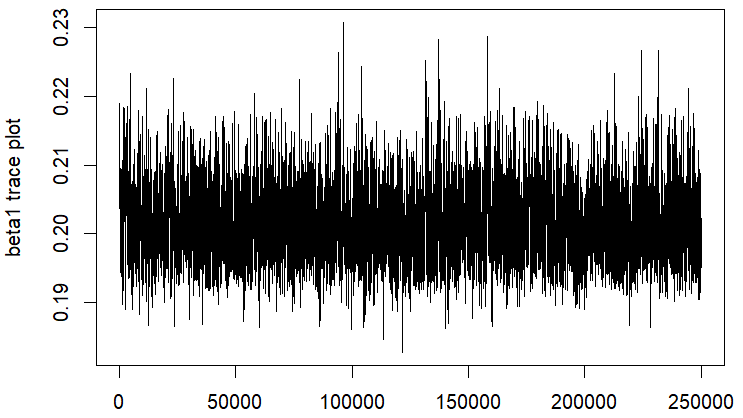} & \includegraphics[width=0.3\textwidth]{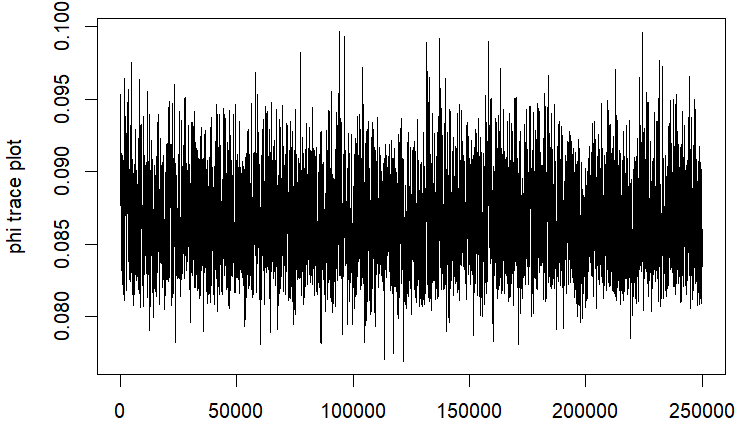}
\end{tabular}
\end{center}
\end{figure}

Figure \ref{fig:braziltraces} shows the trace plots for our main parameters of interest.

\begin{table}[H]
\caption{95\% credible intervals for Brazil Zika outbreak model using increased $J_s$ and $L_s$}
\label{tab:brazil_results_sens}
\begin{center}
\begin{tabular}{ cc|c|c|c }
$J_s$ & $L_s$ & \makecell{$\beta_0$, 95\% \\ cred. interval} & \makecell{$\beta_1$, 95\% \\ cred. interval} & \makecell{$\phi$, 95\% \\ cred. interval} \\
\hline
20 & 10 & (-0.1090, -0.0520) & (0.1933, 0.2128) & (0.0820, 0.0925) \\
25 & 10 & (-0.1086, -0.0535) & (0.1937, 0.2128) & (0.0823, 0.0924) \\
20 & 15 & (-0.1043, -0.0434) & (0.1817, 0.2108) & (0.0817, 0.0919)
\end{tabular}
\end{center}
\end{table}

We next describe some initial work we conducted to fit model-discrepancy terms to the Zika model described in the main text. Specifically, we considered fitting multiplicative model-discrepancy as:
\begin{align}
    y(\bs,t) | \pmb{X}, p, \nu, \bzeta &\sim NB\left(p \lambda(\bs, t), \frac{p \lambda(\bs, t)}{\nu - 1}\right) \\
    \lambda(\bs, t) &:= p \bigl(X(\bs, t - 1) - X(\bs, t)\bigr)\exp\{\mathbf{B}_m \bzeta_{\bs}\}
\end{align}
where $\mathbf{B}_m$ is a matrix of b-splines, $\bzeta = \{\bzeta_1,...,\bzeta_{n_s}\}$, and $\bzeta_{\bs} | \sigma_\zeta^2 \overset{iid}{\sim} N(\pmb{0}, \sigma_\zeta^2 \mathbf{I})$ are coefficients for the multiplicative model discrepancies. We use an $\text{InvGamma}(10,10)$ prior for $\sigma_\zeta^2$.

We experienced convergence problems implementing this model and did not therefore present it in the main text. Nevertheless, we found similar point estimates in our preliminary efforts for $\btheta$, with $\hat{\beta}_0 = -0.107\, (-0.132, -0.084)$, $\hat{\beta}_1 = 0.207\, (0.198, 0.218)$, and $\hat{\phi} = 0.089\, (0.085, 0.094)$, where the numbers in parentheses are the 95\% credible intervals. The following two plots show the new infections for Mato Grosso and Piaui, where the black lines are without the model-discrepancy terms and the red lines are with the model-discrepancy terms (using $\bzeta_s \in \mathbb{R}^5$).

\begin{figure}[H]
\caption{Plots of new infections in Mato Grosso (L) and Piaui (R) with and without model-discrepancy terms (red lines and black lines, respectively).}
\label{fig:multmd}
\begin{center}
\begin{tabular}{cc}
\includegraphics[width=0.45\textwidth]{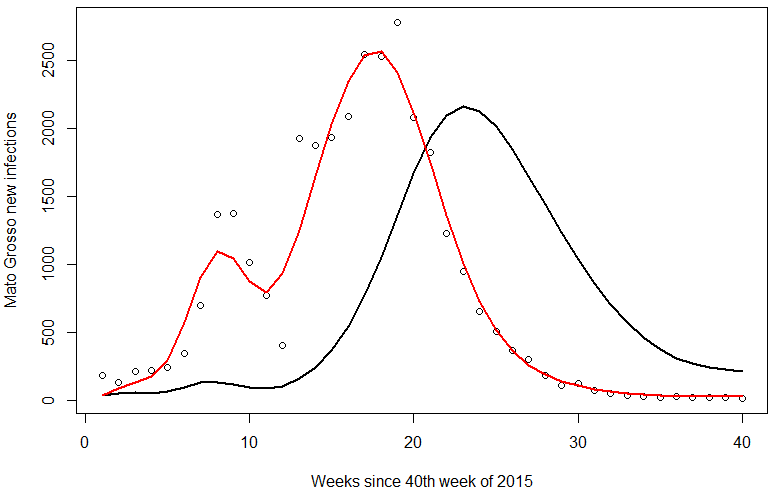} &
\includegraphics[width=0.45\textwidth]{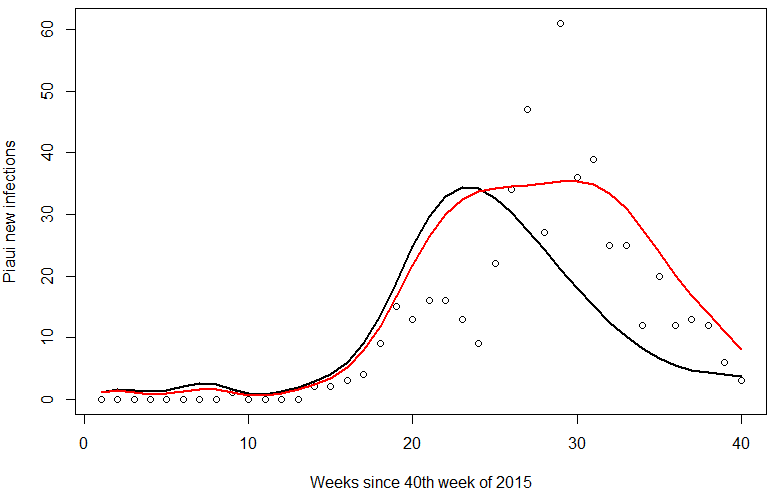}
\end{tabular}
\end{center}
\end{figure}

\end{document}